\documentclass[11pt]{article}
\usepackage[utf8]{inputenc}
\usepackage[T1]{fontenc}
\usepackage{authblk}
\usepackage{amsmath,amssymb,amsthm,mathtools,mathrsfs}
\usepackage{bm,xcolor}
\usepackage{physics}
\usepackage{fullpage}
\usepackage[margin=1in]{geometry}
\usepackage{graphicx,subcaption,epstopdf,float}
\usepackage{booktabs,tabularx,array}
\usepackage{qcircuit}
\usepackage{cite}
\usepackage{enumitem}

\usepackage{tikz-cd}
\usepackage{algorithm}
\usepackage{algpseudocode}
\algrenewcommand{\algorithmicrequire}{\textbf{Input:}}
\algrenewcommand{\algorithmicensure}{\textbf{Output:}}
\usepackage[bookmarksnumbered,colorlinks=true,linkcolor=blue,citecolor=red,urlcolor=blue,hyperindex,linktocpage=true]{hyperref}
\usepackage[nameinlink,capitalise]{cleveref}
\makeatletter
\def\theHALG@line{\arabic{algorithm}.\arabic{ALG@line}}
\makeatother
\usepackage{appendix}

\theoremstyle{definition}
\newtheorem{definition}{Definition}[section]
\newtheorem{problem}{Problem}[section]
\newtheorem{assumption}{Assumption}[section]
\theoremstyle{plain}
\newtheorem{lemma}{Lemma}[section]
\newtheorem{theorem}{Theorem}[section]
\newtheorem*{informaltheorem}{Theorem}
\crefname{assumption}{Assumption}{Assumptions}
\Crefname{assumption}{Assumption}{Assumptions}
\crefname{problem}{Problem}{Problems}
\Crefname{problem}{Problem}{Problems}

\title{Accelerating Diffusion Maps on a Quantum Computer: Spectral Filtering and Galerkin PDE Solvers}

\author[1]{Ruizhe Qian\thanks{rbq5037@psu.edu}}
\author[2]{Xiantao Li\thanks{Xiantao.Li@psu.edu}}

\affil[1]{Department of Mathematics, The Pennsylvania State University, University Park, Pennsylvania 16802, USA}

\date{}

\begin{document}

\maketitle

\begin{abstract}
Diffusion maps generalize principal component analysis (PCA) to nonlinear manifolds, enabling spectral dimensionality reduction, geometric data analysis, and elliptic PDE approximation on point clouds. For large datasets of size $N$, however, classical matrix operations incur prohibitive $\operatorname{poly}(N)$ scaling, even when extracting low-dimensional spectral features or targeted scalar observables. We introduce a quantum algorithmic framework that bypasses explicit graph-Laplacian assembly by exploiting the symmetric factorization $\mathcal{G} = A^T A$ directly at the block-encoding level. Combining Gaussian moment dilation, Laplace-weighted linear combinations of unitaries (LCU), and quantum singular value transformation (QSVT), we implement spectral filters of $A$ to estimate heat-semigroup overlaps, regularized inverse queries, and energy responses, as well as to prepare spectral Galerkin solution states. A common-state construction further allows the block-encoding queries to be used across multiple scalar observables. Under polylogarithmic-cost block-encoding access to $A$ and state-preparation oracles, our heat-overlap algorithm achieves $\mathcal{O}(\operatorname{polylog} N)$ gate complexity at fixed normalized precision, confidence, and spectral scale. Ideal statevector simulations validate our discrete circuit constructions, while a matched approximate-counting benchmark demonstrates the quadratic precision advantage of coherent amplitude estimation for energy readout.
\end{abstract}

\section{Introduction}
\label{sec:introduction}
Diffusion maps provide a geometric framework for nonlinear dimensionality
reduction and spectral data analysis by modeling random walks on a data
similarity graph~\cite{coifman2005geometric,coifman2006diffusion}. Extending
linear methods such as principal component analysis (PCA) to curved data
geometries, density-corrected graph generators converge to the
Laplace--Beltrami operator under appropriate sampling and bandwidth
assumptions, whereas alternative normalizations retain sampling-density
information~\cite{nadler2006diffusion}. These spectral properties have made
diffusion maps a foundational tool across scientific computing and data
science, with applications ranging from reduced-order modeling of stochastic
dynamical systems~\cite{coifman2008reduction} to single-cell developmental
trajectories~\cite{haghverdi2015diffusion,angerer2016destiny}, cortical
connectivity gradients~\cite{margulies2016situating}, and function
approximation and partial differential equations (PDEs) on sampled manifolds.

As dataset sizes grow, however, constructing and manipulating the resulting
kernel matrices and discrete differential operators incurs substantial
$\operatorname{poly}(N)$ computational overhead, even when downstream tasks
require only targeted scalar observables or low-dimensional projections.
This bottleneck motivates the use of quantum algorithms capable of extracting
selected spectral quantities with provable scaling advantages. Advances in
quantum linear-system solvers and quantum singular value transformation
(QSVT) establish that matrix functions can be evaluated with polylogarithmic
dependence on the state-space dimension $N$, provided coherent access,
conditioning, and precision requirements are
met~\cite{harrow2009quantum,childs2017quantum,gilyen2019quantum}. Moreover,
recent frameworks for quantum processing of massive classical data demonstrate
that structured or sparse data matrices can be queried via block-encodings
without prohibitive memory overhead~\cite{zhao2026exponential}, creating a
natural pathway for quantum spectral methods on large point clouds.

Building on quantum PDE discretization techniques~\cite{li2026quantum}, we
develop a unified quantum algorithmic framework for diffusion maps and
manifold elliptic equations centered on the symmetric factorization
$\mathcal G = A^T A$, where $A$ denotes a weighted discrete gradient or
energy factor. Rather than assembling the graph Laplacian $\mathcal G$
explicitly, our approach operates directly on a block-encoding of $A$ to
address three core computational tasks: estimating heat-semigroup matrix
elements, evaluating regularized inverse quantities, and preparing spectral
Galerkin solution states along with their associated energy responses. By
combining QSVT~\cite{gilyen2019quantum}, Gaussian moment
dilation~\cite{wu2026quantum}, and coherent observable
estimation~\cite{rall2020quantum}, all three tasks share a single factor
oracle and common-state construction. A detailed comparison with prior
literature is provided in \cref{sec:related-work}.

Throughout this work, $N$ denotes the number of point-cloud samples on a
manifold $\mathcal M \subset \mathbb R^D$ of ambient dimension $D$ and
intrinsic dimension $d = \dim \mathcal M$ (or, equivalently, the number of
degrees of freedom in a PDE discretization), and our complexity comparisons
focus on scaling with $N$. For the continuous-time heat operator
$K(t) = e^{-t\mathcal G}$ at diffusion time $t \geq 0$, let
$\alpha_A \geq \|A\|$ be the block-encoding normalization factor in
\cref{ass:matrix-access}. Under polylogarithmic-cost coherent access, bounded
$\alpha_A\sqrt{t}$, and fixed normalized-overlap accuracy and confidence, our
heat-kernel estimator achieves polylogarithmic gate complexity in $N$. By
contrast, generic sparse explicit-vector methods scale linearly with $N$, as
summarized in \cref{tab:heat-dimension-comparison}. We emphasize that this
separation compares against full-vector sparse solvers rather than establishing
an unconditional classical lower bound, as local classical diffusion
routines and random-walk estimators can also bypass full-graph traversal for
certain localized queries~\cite{kloster2014heat,linden2022heat}; furthermore,
end-to-end application accuracy must account for output normalization factors.

For elliptic PDEs on manifolds, we instead seek the solution projected onto a
low-frequency spectral subspace. Rather than explicitly computing an
eigenbasis or assembling a reduced Galerkin stiffness matrix, our algorithms
filter the spectrum directly via the factor oracle $A$ to either prepare the
corresponding quantum solution state or estimate its quadratic energy
response. Under bounded spectral scales, source norm, and target accuracy,
together with polylogarithmic-cost coherent access, both tasks retain
polylogarithmic gate complexity in $N$ (with state preparation additionally
requiring a nonvanishing source-overlap lower bound, whereas direct energy
estimation avoids any such overlap assumption). Beyond dimension scaling,
coherent readout also yields a quadratic precision advantage: in
\cref{sec:numerical-query-advantage}, we construct a matched bit-oracle
benchmark whose energy query reduces to approximate counting, establishing
worst-case query complexities of $\Theta(\epsilon_E^{-1})$ quantumly versus
$\Theta(\epsilon_E^{-2})$ classically at fixed confidence in the regime
$N \gg \epsilon_E^{-2}$~\cite{brassard2002quantum,nayak1999quantum}.

\subsection{Problem setup}

We write classical vectors in bold, such as $\bm{x}$, and quantum states
in Dirac notation, such as $\ket{x}$. For $\bm{x}\neq0$, the normalized
vector is $\hat{\bm{x}}=\bm{x}/\|\bm{x}\|$, and $\ket{\hat{x}}$ denotes
its amplitude-encoded quantum state.

Let $\bm{x}_1,\ldots,\bm{x}_N\in\mathcal M\subset\mathbb R^D$ be data samples,
with symmetric affinities $w_{ij}\geq0$, $w_{ii}=0$, and an edge whenever
$w_{ij}>0$. Write $W=(w_{ij})$, $d_i=\sum_jw_{ij}>0$, and
$D=\operatorname{diag}(d_1,\ldots,d_N)$. The diffusion-map transition
matrix $P=D^{-1}W$ and its symmetric conjugate $S$ satisfy
\[
S=D^{1/2}PD^{-1/2}=D^{-1/2}WD^{-1/2},\qquad
\mathcal G=I-S=D^{-1/2}(D-W)D^{-1/2}\succeq0.
\]
Here we follow the standard diffusion-map notation $\mathcal G$ for the
symmetric normalized graph Laplacian; in particular, $\mathcal G$ is a
Hermitian matrix operator on $\mathbb R^N$ and should not be confused with
a quantum channel or superoperator. Classical diffusion maps use powers of
$P$~\cite{coifman2005geometric,coifman2006diffusion}. We study their
continuous-time counterpart,
$e^{-t(I-P)}=D^{-1/2}e^{-t\mathcal G}D^{1/2}$.

\begin{definition}[Continuous-time graph diffusion kernel]
\label{def:diffusion-kernel}
For $t\geq0$, the diffusion kernel in symmetric coordinates is
\begin{equation}
\label{eq:diffusion-kernel}
K(t):=e^{-t\mathcal G}.
\end{equation}
Its entries are $K_{ij}(t)=\bm{e}_i^TK(t)\bm{e}_j$, where $\bm{e}_i$ is the $i$th
standard basis vector.
\end{definition}

Orient the $M$ edges arbitrarily and let $C\in\mathbb R^{M\times N}$
have entries $\sqrt{w_{ij}}$ and $-\sqrt{w_{ij}}$ in the row for edge
$(i,j)$, with zeros elsewhere. Then
\[
C^TC=D-W,\qquad A=CD^{-1/2},\qquad
\mathcal G=A^TA,\qquad \bm{v}^T\mathcal G\bm{v}=\|A\bm{v}\|^2.
\]
Thus the normalized discrete gradient $A\in\mathbb R^{M\times N}$ supplies
the energy factor used by our algorithms. 

From a quantum algorithmic standpoint, assuming coherent block-encoding
access to $A$ is well motivated. Recent advances in quantum data
processing demonstrate that sparse-access oracles and block-encodings for
massive classical matrices can be instantiated directly from classical data
streams or samples—for example, via quantum oracle sketching—without
requiring large-scale quantum random-access memory
(QRAM)~\cite{zhao2026exponential}. Because each row of $A$ corresponds to a
single graph edge $(i,j)$ with at most two nonzero entries determined by
local edge weights and vertex degrees, $A$ is $2$-row-sparse and directly
compatible with sparse block-encoding primitives, whereas assembling
$\mathcal{G}=A^TA$ explicitly would incur quadratic overhead in the
dataset size.

\paragraph{From Gaussian affinities to a manifold generator.}
To connect discrete graph diffusion to continuous differential operators
on manifolds, consider the normalized Gaussian integral kernel on a smooth
compact embedded manifold $\mathcal M$ without
boundary~\cite{coifman2006diffusion,berry2016variable},
\[
k_\sigma(\bm{x},\bm{y})=\exp\!\left(-\frac{\|\bm{x}-\bm{y}\|^2}{4\sigma^2}\right),\qquad
(T_\sigma\varphi)(\bm{x})=
\frac{\int_{\mathcal M}k_\sigma(\bm{x},\bm{y})\varphi(\bm{y})\,d\mathrm{vol}_{\mathcal M}(\bm{y})}
{\int_{\mathcal M}k_\sigma(\bm{x},\bm{y})\,d\mathrm{vol}_{\mathcal M}(\bm{y})}.
\]
For sufficiently smooth $\varphi$,
$T_\sigma\varphi=\varphi+\sigma^2\Delta_{\mathcal M}\varphi+O(\sigma^4)$.
Here $\Delta_{\mathcal M}=\operatorname{div}_{\mathcal M}\nabla_{\mathcal M}$
has nonpositive spectrum, so $(I-T_\sigma)/\sigma^2$ approximates
$-\Delta_{\mathcal M}$. The corresponding discrete scaling is
\[
L_\sigma=\sigma^{-2}\mathcal G_\sigma
=(\sigma^{-1}A_\sigma)^T(\sigma^{-1}A_\sigma),\qquad
e^{-\tau L_\sigma}=e^{-t\mathcal G_\sigma},\qquad t=\tau/\sigma^2.
\]
This continuum limit justifies using graph spectral filters to solve
elliptic PDEs on $\mathcal M$, with physical diffusion time $\tau$ and
kernel bandwidth $\sigma$ entering solely through the dimensionless graph
diffusion time $t=\tau/\sigma^2$. Because $\sigma$ can always be absorbed
into $t$ (or into a constant rescaling of the factor $A$), we suppress the
subscript $\sigma$ henceforth and work directly with $\mathcal G=A^TA$ and
dimensionless parameter $t$. More generally, $A$ may factor any discrete
elliptic energy; this common structure supports the following three tasks.

\subsubsection{Heat-semigroup matrix-element estimation}
\label{sec:setup-heat}

Classical exponential-action methods compute a diffused signal $K(t)\bm{h}$
using polynomial or Krylov approximations~\cite{hochbruck1997krylov,almohy2011action}.
Selected predictions and aggregate queries instead require matrix elements.
For $\bm{h}\neq0$, with $|\hat h\rangle=\bm{h}/\|\bm{h}\|$,
\[
(K(t)\bm{h})_i=\|\bm{h}\|\langle e_i|K(t)|\hat h\rangle.
\]

\begin{problem}[Heat-semigroup matrix-element estimation]
\label{prob:heat}
Given $t\geq0$, normalized real states $|u\rangle,|v\rangle\in\mathbb R^N$,
an additive tolerance $\epsilon>0$, and failure tolerance
$0<\delta_{\mathrm{fail}}<1/2$, estimate
$z(t):=\langle u|K(t)|v\rangle$ to error $\epsilon$ with probability
at least $1-\delta_{\mathrm{fail}}$.
For $J\geq1$ normalized real query states sharing $|v\rangle$, estimate
\[
\begin{gathered}
m_i(t):=\langle u_i|K(t)|v\rangle,\quad i=1,\ldots,J,\\
\Pr\!\left[\max_i|\widehat m_i(t)-m_i(t)|\leq\epsilon\right]
\geq1-\delta_{\mathrm{fail}}.
\end{gathered}
\]
\end{problem}

Throughout the complexity summaries, $\widetilde O$ suppresses logarithmic
factors in accuracy, confidence, and the relevant normalized spectral parameters.
Query counts include inverse and required controlled calls.

\begin{informaltheorem}[Heat-estimation complexity]
Under the access assumptions of
\cref{thm:single-heat-estimation,thm:joint-heat-estimation},
the matrix-oracle counts for one overlap and $J$ overlaps are, respectively,
\[
\widetilde O\!\left(\frac{1+\alpha_A\sqrt t}{\epsilon}\right),
\qquad
\widetilde O\!\left(\frac{\sqrt J(1+\alpha_A\sqrt t)}{\epsilon}\right).
\]
A single overlap requires $\widetilde O(1/\epsilon)$ state-preparation calls.
The joint bound counts shared operator work, with observable-specific
operations and the $J$ classical outputs charged separately.
\end{informaltheorem}

For comparison, independent Hadamard sampling, independent coherent
estimation, and joint coherent estimation use
$\widetilde O(J/\epsilon^2)$, $\widetilde O(J/\epsilon)$, and
$\widetilde O(\sqrt J/\epsilon)$ calls to the shared heat block,
respectively. The joint procedure combines our common-state construction
with the estimator of Huggins et al.~\cite{huggins2022multiple}.

\subsubsection{Regularized inverse scalar estimation}
\label{sec:setup-inverse}

While the heat semigroup $e^{-t\mathcal G}$ captures transient diffusion
over a finite timescale $t$, many inference and inverse problems on point
clouds instead call for steady-state or globally regularized responses
governed by the resolvent $H_\beta^{-1}:=(\mathcal G+\beta I)^{-1}$ with
$\beta>0$. A canonical example arises in manifold regularization and
semi-supervised learning, where noisy or partial observations $\bm{\ell}\neq0$
must be interpolated smoothly across the underlying data
geometry~\cite{zhou2003consistency,belkin2006manifold}. Balancing
geometric smoothness, measured by the discrete Dirichlet energy
$\bm{z}^T\mathcal G\bm{z}=\|A\bm{z}\|^2$, against fidelity to $\bm{\ell}$ leads to the
graph-regularized least-squares problem
\[
\bm{z}_\beta=\arg\min_{\bm{z}\in\mathbb R^N}
\left\{\frac12\bm{z}^T\mathcal G\bm{z}+\frac\beta2\|\bm{z}-\bm{\ell}\|^2\right\}
=\beta H_\beta^{-1}\bm{\ell}.
\]
The minimizer is unique because $H_\beta\succeq\beta I$.
For a normalized graph Laplacian and initial class scores that vanish at
unlabeled nodes, this is graph label propagation~\cite{zhou2003consistency}.
A selected score is
$(\bm{z}_\beta)_i=\beta\|\bm{\ell}\|\langle e_i|H_\beta^{-1}|\hat\ell\rangle$,
where $|\hat\ell\rangle=\bm{\ell}/\|\bm{\ell}\|$.

\begin{problem}[Regularized inverse scalar estimation]
\label{prob:inverse}
Given $\beta>0$, normalized real states $|y\rangle,|b\rangle\in\mathbb R^N$,
$\epsilon>0$, and $0<\delta_{\mathrm{fail}}<1/2$, estimate
\[
r_\beta(\bm{y},\bm{b}):=\bm{y}^T\bm{x}_\beta=\langle y|H_\beta^{-1}|b\rangle,
\qquad \bm{x}_\beta=H_\beta^{-1}\bm{b},
\]
to additive error $\epsilon$ with probability at least
$1-\delta_{\mathrm{fail}}$.
\end{problem}

Classically, evaluating $r_\beta(\bm{y},\bm{b})$ requires solving the shifted
Laplacian system $H_\beta \bm{x}_\beta=\bm{b}$---for instance, via conjugate
gradients---and computing $\bm{y}^T\bm{x}_\beta$, or approximating the inverse
bilinear form via Lanczos-based Gauss
quadrature~\cite{li2016quadrature}.

\begin{informaltheorem}[Regularized-inverse estimation complexity]
Under \cref{thm:inverse-scalar-estimation}, the scalar query requires
\[
\widetilde O\!\left(\frac{1+\alpha_A/\sqrt\beta}{\beta\epsilon}\right)
\]
matrix-oracle calls and $\widetilde O(1/(\beta\epsilon))$
state-preparation calls when $\beta\epsilon<1$.
If $\beta\epsilon\geq1$, returning zero suffices.
\end{informaltheorem}

\subsubsection{Spectral Galerkin solutions and energy responses}
\label{sec:setup-galerkin}

Elliptic partial differential equations on sampled manifolds arise across
scientific and engineering applications—from surface diffusion and
equilibrium transport on curved interfaces to committor and mean
first-passage time equations on high-dimensional dynamical manifolds—where
constructing a volumetric mesh is infeasible. In this meshfree setting,
low-frequency graph Laplacian modes provide natural continuum-consistent
trial spaces while filtering out high-frequency point-cloud sampling
noise~\cite{yan2023spectral}. Specifically, we consider the variable-coefficient
elliptic equation
\[
\mathcal L u=-\frac1\rho\operatorname{div}_{\mathcal M}
(\rho\,\kappa\nabla_{\mathcal M}u)+cu=g
\]
on a smooth compact manifold $\mathcal M\subset\mathbb R^D$ without boundary,
with smooth weight $\rho>0$, symmetric uniformly positive definite
conductivity tensor $\kappa$, and reaction coefficient $c\geq0$. After
discretization with diagonal mass matrix $M_N=\operatorname{diag}(m_i)$ and
change to mass-weighted coordinates $\bm{u}=M_N^{1/2}\bm{z}$ (so that the discrete
$L^2(\rho\,d\mathrm{vol}_{\mathcal M})$ inner product becomes Euclidean),
suppose the discrete operator factors as $L=\mathcal G=A^TA$ with
right-hand side $\bm{f}=M_N^{1/2}\bm{g}_h$.

To capture macroscopic behavior while regularizing high-frequency noise, we
restrict the solution to the low-frequency eigenspace of $L$ below a
spectral cutoff $\delta>0$. Define the spectral projectors
\[
P_0=\mathbf1_{\{0\}}(L),\quad
\Pi_\delta=\mathbf1_{[0,\delta]}(L),\quad
P_\delta^+=\Pi_\delta-P_0,\quad
\mathcal E_\delta=\operatorname{Ran}(P_\delta^+).
\]
For $\bm{f}\neq0$ and quadratic energy functional $J(\bm{v})=\frac12\|A\bm{v}\|^2-\bm{f}^T\bm{v}$,
the Galerkin minimizer and its associated energy response are given by
\[
\bm{u}_\delta=\arg\min_{\bm{v}\in\mathcal E_\delta}J(\bm{v})=L^+P_\delta^+\bm{f},
\qquad
E_\delta=\bm{f}^T\bm{u}_\delta=\|A\bm{u}_\delta\|^2=-2J(\bm{u}_\delta),
\]
where $L^+$ denotes the Moore--Penrose pseudoinverse. Equivalently,
$L\bm{u}_\delta=P_\delta^+\bm{f}$ with $\bm{u}_\delta\in\mathcal E_\delta$; projecting out
the nullspace $P_0$ guarantees a unique solution for arbitrary $\bm{f}$.
Writing $B_\delta=\Pi_\delta L\Pi_\delta$ for the truncated operator, we
have $B_\delta^+=L^+P_\delta^+$ and $\bm{u}_\delta=B_\delta^+\bm{f}$. In physical
and engineering models, $E_\delta$ directly quantifies macroscopic
quantities such as Dirichlet energy, compliance, steady-state dissipation,
or integrated linear response functionals.

\begin{problem}[Spectral Galerkin state preparation and energy estimation]
\label{prob:galerkin}
Given $L=A^TA$, $\delta>0$, $\bm{f}\neq0$, and
$0<\delta_{\mathrm{fail}}<1/2$, consider two outputs.

\emph{Solution state.} If $P_\delta^+\bm{f}\neq0$, prepare
$|\hat u_\delta\rangle=\bm{u}_\delta/\|\bm{u}_\delta\|$ to accuracy
$0<\epsilon_{\mathrm{st}}<1$. Return a success flag with probability
at least $1-\delta_{\mathrm{fail}}$ and, conditional on success,
a normalized state $|\psi_\delta\rangle$ satisfying
\[
\min_{\vartheta\in\mathbb R}
\bigl\||\psi_\delta\rangle-e^{i\vartheta}|\hat u_\delta\rangle\bigr\|_2
\leq\epsilon_{\mathrm{st}}.
\]

\emph{Energy response.} Given $F=\|\bm{f}\|$ and $\epsilon_E>0$, return
$\widehat E_\delta\in\mathbb R$ with
$\Pr[|\widehat E_\delta-E_\delta|\leq\epsilon_E]\geq1-\delta_{\mathrm{fail}}$.
This includes $P_\delta^+\bm{f}=0$, for which $E_\delta=0$.
\end{problem}

In physical nodal coordinates $\bm{z}_\delta=M_N^{-1/2}\bm{u}_\delta$, the quantum
state $|\hat u_\delta\rangle$ encodes amplitudes proportional to
$\sqrt{m_i}(\bm{z}_\delta)_i$. Computational-basis measurements thus sample
mass-weighted squared solution values, whereas signed nodal or regional
averages can be extracted via scalar overlap estimation. Unlike classical
spectral Galerkin solvers, which must explicitly compute the leading
eigenvectors of $L$ and assemble a reduced stiffness matrix, our quantum
algorithms apply the filtered pseudoinverse $L^+P_\delta^+$ directly via
the factor oracle $A$. When an upper spectral cutoff is implemented via
polynomial filtering across a transition window $[\delta_-,\delta_+]$
containing $\delta$, the circuit complexity depends on the singular-value
separation $\Delta_\sigma=\sqrt{\delta_+}-\sqrt{\delta_-}>0$.

\begin{informaltheorem}[Spectral Galerkin complexity]
Under \cref{ass:matrix-access,ass:source-oracle,ass:galerkin-spectrum},
let $\gamma>0$ bound the positive eigenvalues of $L$ from below.
When an upper cutoff is imposed, let
$\Delta_\sigma=\sqrt{\delta_+}-\sqrt{\delta_-}>0$ denote its singular-value
separation. The procedures in
\cref{thm:galerkin-state,alg:galerkin-energy} achieve the above guarantees with
matrix-oracle counts
\[
\begin{aligned}
\text{Solution state:}\quad &
\widetilde O\!\left(\frac1{s_0}
\left[\frac{\alpha_A}{\sqrt\gamma}+\frac{\alpha_A}{\Delta_\sigma}\right]\right),\\
\text{Energy response:}\quad &
\widetilde O\!\left(\frac{F^2}{\gamma\epsilon_E}
\left[\frac{\alpha_A}{\sqrt\gamma}+\frac{\alpha_A}{\Delta_\sigma}\right]\right).
\end{aligned}
\]
State preparation requires a known bound
$0<s_0\leq\gamma\|\bm{u}_\delta\|/(4F)$ and $\widetilde O(1/s_0)$ source calls.
Its suppressed logarithms include $1/(\epsilon_{\mathrm{st}}s_0)$.
Energy estimation requires $\widetilde O(F^2/(\gamma\epsilon_E))$ source
calls and no overlap lower bound. If $\epsilon_E\geq F^2/\gamma$,
returning zero suffices. When all positive modes are retained, omit
the terms $\alpha_A/\Delta_\sigma$.
\end{informaltheorem}

Throughout, $\delta$ denotes the spectral cutoff and $\delta_{\mathrm{fail}}$
the failure tolerance, while $\epsilon$, $\epsilon_E$, and
$\epsilon_{\mathrm{st}}$ denote the target accuracies for heat or inverse
scalar queries, energy responses, and normalized solution states,
respectively.

\subsection{Quantum Oracle Access}
\label{sec:quantum-oracle-access}

\begin{definition}[Block-encoding~\cite{gilyen2019quantum}]
\label{def:block-encoding}
For $X\in\mathbb C^{n\times n}$, $\alpha\geq\|X\|$ with $\alpha>0$,
$a\in\mathbb N_0$, and $\eta\geq0$, an $(\alpha,a,\eta)$-block-encoding
is a unitary $U\in\mathbb C^{2^an\times2^an}$ satisfying
\[
\bigl\|X-\alpha(\bra0^{\otimes a}\otimes I)U
(\ket0^{\otimes a}\otimes I)\bigr\|\leq\eta.
\]
\end{definition}

\begin{assumption}[Coherent access to the energy factor]
\label{ass:matrix-access}
For $A\in\mathbb R^{M\times N}$, let $\overline A$ be its zero padding
to a square matrix of power-of-two dimension. We are given a known
$\alpha_A\geq\|A\|>0$ and an exact $(\alpha_A,a_A,0)$-block-encoding
$U_A$ of $\overline A$. The circuits $U_A$, $U_A^\dagger$, and their
required controlled versions have gate cost $T_A$ per application,
including control overhead.
\end{assumption}

Factor access also supplies the Hermitian dilation through
\[
B_A=\begin{pmatrix}0&A^T\\A&0\end{pmatrix},\qquad
U_B=|0\rangle\langle1|\otimes U_A^\dagger
+|1\rangle\langle0|\otimes U_A.
\]
The Hermitian unitary $U_B$ is an exact $(\alpha_A,a_A,0)$-block-encoding
of the padded $B_A$, with $\|B_A\|=\|A\|$ and the sector qubit in the
system register. Composing $U_A^\dagger$ and $U_A$ with separate
block-ancilla registers gives an exact $(\alpha_A^2,2a_A,0)$-block-encoding
of the padded $\mathcal G=A^TA$~\cite{gilyen2019quantum}.
Both derived oracles and their required controls cost $O(T_A)$.
We suppress zero-padding notation below.

\begin{assumption}[State preparation]
\label{ass:state-prep}
For each normalized real input $|\psi\rangle\in\mathbb R^N$,
a unitary $U_\psi$ prepares $U_\psi|0\rangle=|\psi\rangle$ at gate cost
$T_\psi$. Its inverse and required controlled versions cost $O(T_\psi)$.
\end{assumption}

\begin{assumption}[Source Oracle]
\label{ass:source-oracle}
A known reversible unitary $U_f$ prepares
$U_f|0\rangle=|\hat f\rangle=\bm{f}/F$, where $F=\|\bm{f}\|>0$.
It, its inverse, and required controlled versions are available at
cost $T_f$ per call. Energy estimation also receives $F$ as a classical input.
\end{assumption}

These assumptions concern a supplied discrete problem. Graph construction,
degree computation, mass normalization, and input-state preparation enter
through preprocessing and oracle costs. Approximate oracles require
corresponding error accounting within the complete circuits.

\subsection{Related Work}
\label{sec:related-work}

Quantum algorithms for nonlinear manifold learning have largely focused on
reconstructing low-dimensional coordinate embeddings or geodesic distances
from point-cloud data. In the context of diffusion maps, Sornsaeng
et al.~\cite{sornsaeng2021quantum} construct a quantum Markov transition
operator, estimate its leading eigenpairs via phase estimation, and recover
classical coordinates through eigenvector tomography, while Nghiem
et al.~\cite{nghiem2025intrinsic} develop quantum routines to estimate
local geometric quantities and diffusion coordinates of individual target
points. Related dimensionality-reduction frameworks follow a similar
embedding-centric paradigm, including quantum locally linear embedding via
linear-algebraic and variational solvers~\cite{he2020quantum} and quantum
Isomap via quantum shortest-path search and multidimensional
scaling~\cite{feng2024quantum}. Taking a complementary dynamical perspective,
Kumar and Sarovar~\cite{kumar2022shining} employ quantum wavepacket
propagation on graph Laplacians to probe geodesic geometry and manifold
structure. Our work shares this operator-dynamics viewpoint but shifts the
computational target from explicit coordinate recovery to direct spectral
filtering: given coherent access to the factor $A$, we evaluate
heat-semigroup overlaps and regularized inverse bilinear forms of
$\mathcal G = A^T A$, and prepare low-frequency elliptic solution states,
entirely bypassing eigenbasis reconstruction and eigenvector tomography.

These spectral observables are rooted in classical manifold learning and
point-cloud PDE discretizations. Whereas diffusion-map kernel ridge
regression regularizes a dense kernel Gram matrix~\cite{harlim2026diffusion},
our inverse queries regularize the graph generator directly, aligning with
graph-based semi-supervised learning and label
propagation~\cite{zhou2003consistency}. In parallel, classical weak-form and
spectral Galerkin methods discretize elliptic equations on sampled
manifolds using low-lying Laplacian or operator
modes~\cite{vaughn2019diffusion,yan2023spectral}. By formulating the
Galerkin problem directly in the eigenbasis of the discrete elliptic
operator, our framework enables direct quantum filtering of its projected
pseudoinverse without assembling a reduced stiffness matrix.

Algorithmically, our spectral-filtering routines connect to quantum
Markov-chain fast-forwarding~\cite{apers2019fast}, local heat
estimation~\cite{linden2022heat}, matrix-function entry
estimation~\cite{montanaro2024functions}, and factored-generator techniques
such as Gaussian moment dilation~\cite{wu2026quantum},
transmutation~\cite{jin2026transmutation}, and multilevel parabolic
solvers~\cite{li2026parabolic} (as well as Hermitian-lift representations in
quantum Gibbs sampling~\cite{chowdhury2016quantum}). In particular,
combining our factor oracle $A$ with a Laplace-weighted unitary integral
representation---building on the Lap-LCHS framework of An
et al.~\cite{an2026laplace}---implements the regularized resolvent
$\beta(\mathcal G+\beta I)^{-1}$ with operator query complexity
$\widetilde O(1+\alpha_A/\sqrt\beta)$, achieving a quadratic conditioning
improvement over the $\widetilde O(1+\alpha_A^2/\beta)$ scaling of direct
Lap-LCHS applied to a product block-encoding of $\mathcal G$. Quantum
singular value transformation (QSVT) provides a complementary polynomial
pathway for implementing factor-based heat and inverse
filters~\cite{gilyen2019quantum,li2026parabolic}.

For solution state preparation and scalar observable extraction, our
methods draw on quantum linear-system and QSVT
primitives~\cite{harrow2009quantum,childs2017quantum,gilyen2019quantum,chakraborty2023regularized},
quantum finite-element functional
analysis~\cite{montanaro2016finite,deiml2026quantum}, and coherent
observable estimation~\cite{rall2020quantum,huggins2022multiple}, where our
common-state architecture explicitly separates shared spectral-filter calls
from observable-specific operations and classical readout costs. On the
input side, recent constructions of weighted-Laplacian block encodings from
quantum random access~\cite{liu2023laplacian} and streaming matrix-sketching
primitives~\cite{zhao2026exponential} demonstrate how structured graph
operators can be interfaced with quantum processors. Adopting the controlled factor and state-preparation oracles of Assumptions 1.1–1.3 as our computational boundary cleanly decouples the query complexity of spectral filtering from upstream graph construction and data loading.

The remainder of the paper is organized as follows.
\Cref{sec:algorithms-and-theoretical-analysis} presents our quantum
algorithms and theoretical guarantees, \cref{sec:numerical-discussions}
reports numerical validations and precision benchmarks, and
\cref{sec:summary} concludes with a summary and outlook. Technical proofs and
circuit-level implementation details are deferred to \cref{sec:appendix}.

\section{Algorithms and Theoretical Analysis}
\label{sec:algorithms-and-theoretical-analysis}
\subsection{Quantum Primitives for Operator Construction and Estimation}
\label{sec:quantum-primitives}

We implement functions of $\mathcal{G}=A^TA$ in
\cref{prob:heat,prob:inverse,prob:galerkin} using quantum singular value transformation
(QSVT)~\cite{gilyen2019quantum} and Gaussian moment
dilation~\cite{kharazi2026sublinear,wu2026quantum,jin2026transmutation}.
Both use the Hermitian dilation $B_A$ from \cref{ass:matrix-access}, with no separate
oracle for $\mathcal G$.
Rall's observable-estimation algorithm~\cite{rall2020quantum} then extracts scalar outputs.
Proofs of these primitives are given in the cited papers. Our appendix
records their specialization to the present operators and readout tasks.

The problem formulations use real data matrices, sources, and queries,
embedded in complex Hilbert spaces for quantum implementation. We use $T$ for transpose and $\dagger$ for complex adjoint,
which coincide for real matrices. Bras denote conjugate transposes.
Circuits and intermediate states may have complex amplitudes. Spaces are padded
to qubit registers when needed.

Let $E_N:\mathbb{C}^N\rightarrow\mathbb{C}^N\oplus\mathbb{C}^M$ be the real isometry
$E_N\bm{v}=\binom{\bm{v}}{\bm{0}}$ (or $E_N= \ket{0} \otimes I$), with $E_N^{\dagger}=E_N^T$. We pad both summands to a common
power-of-two dimension at least $\max\{N,M\}$ and use a tag qubit to identify the
original system sector. Then
\begin{equation}
\label{eq:dilation-square}
B_A=\begin{pmatrix}0&A^T\\A&0\end{pmatrix}, \qquad
B_A^2=\begin{pmatrix}\mathcal{G}&0\\0&AA^T\end{pmatrix}, \qquad
E_N^{\dagger}B_A^2E_N=\mathcal{G}.
\end{equation}
The key observation here is that compressing even functions of $B_A$ to the original system space therefore gives
functions of $\mathcal{G}$, as in the factored-generator construction
of~\cite[Section~5.3]{wu2026quantum}.

\subsubsection{Quantum singular value transformation}

QSVT provides the spectral filters and inverse transformations used to prepare the
spectral Galerkin solution in \cref{prob:galerkin}.

\begin{definition}[Quantum singular value transformation]
\label{def:qsvt}
Let $A\in\mathbb{R}^{M\times N}$ have rank $r$ and singular value decomposition
$A=\sum_{j=1}^r\sigma_j \bm{u}_j\bm{v}_j^T$, where $\sigma_j>0$. Complete $\{\bm{v}_j\}_{j=1}^r$ to an
orthonormal basis of $\mathbb{R}^N$ and set $\sigma_j=0$ for $j>r$. For a real polynomial $p$
of degree $d$ satisfying
\begin{equation}
\label{eq:qsvt-admissibility}
p(-x)=(-1)^d p(x), \qquad |p(x)|\leq1 \quad\text{for all }x\in[-1,1],
\end{equation}
the singular value transformation of $A/\alpha_A$ is
\begin{equation}
\label{eq:singular-value-transform}
p^{(\mathrm{SV})}(A/\alpha_A)=
\begin{cases}
\displaystyle\sum_{j=1}^r p(\sigma_j/\alpha_A)\bm{u}_j\bm{v}_j^T, & d\text{ odd},\\[6pt]
\displaystyle\sum_{j=1}^N p(\sigma_j/\alpha_A)\bm{v}_j\bm{v}_j^T, & d\text{ even}.
\end{cases}
\end{equation}
Real-polynomial QSVT~\cite[Corollary~18 in the full version]{gilyen2019quantum}
implements this transformation with unit block normalization, $O(d)$ calls to the
input block encoding and its adjoint, signal-subspace phase rotations, and one
additional signal ancilla. The implementation tolerance includes phase-computation
and gate-synthesis errors.
\end{definition}

For even $p$, the dilation oracle gives
\begin{equation*}
p^{(\mathrm{SV})}(A/\alpha_A)=E_N^{\dagger}p(B_A/\alpha_A)E_N.
\end{equation*}
This compression includes the action on the nullspace of $\mathcal{G}$.

\subsubsection{Gaussian moment dilation}

Gaussian moment dilation~\cite{kharazi2026sublinear,wu2026quantum,jin2026transmutation}
expresses the heat operator as an average of unitary propagators. It supplies the
operator in \cref{prob:heat} and, through a Laplace transform, the regularized inverse
in \cref{prob:inverse}.

\begin{definition}[Gaussian moment dilation]
\label{def:gaussian-dilation}
Let $B$ be a finite-dimensional Hermitian operator. On
$\mathcal{H}_{\mathrm{aux}}=L^2(\mathbb{R};\mathbb C)$, let $Q$ be the self-adjoint multiplication
operator $(Qh)(q)=qh(q)$ on its natural domain, and let $\ket{g}$ be the normalized state with
wavefunction
\begin{equation}
g(q)=(4\pi)^{-1/4}e^{-q^2/8}, \qquad
|g(q)|^2=\frac{1}{\sqrt{4\pi}}e^{-q^2/4}.
\end{equation}
For $t\geq0$, the unitary
\begin{equation}
\label{eq:gaussian-unitary}
V_B(t)=e^{-i\sqrt{t}\,Q\otimes B}
\end{equation}
is a Gaussian moment dilation of $e^{-tB^2}$, with compression
\begin{equation}
\label{eq:gaussian-compression}
(\bra{g}\otimes I)V_B(t)(\ket{g}\otimes I)=e^{-tB^2}.
\end{equation}
\end{definition}

The Gaussian has variance $2$ and moments
\begin{equation}
\label{eq:gaussian-moments}
\bra{g}Q^{2m+1}\ket{g}=0,\qquad
\bra{g}Q^{2m}\ket{g}=\frac{(2m)!}{m!},\qquad m\geq0.
\end{equation}

Equivalently, the compression is the Gaussian average
\begin{equation}
\label{eq:gaussian-integral}
e^{-tB^2}=\frac{1}{\sqrt{4\pi}}\int_{\mathbb{R}}e^{-q^2/4}
 e^{-i\sqrt{t}\,qB}\,dq.
\end{equation}
Applying \eqref{eq:dilation-square} gives the diffusion kernel in \cref{def:diffusion-kernel}:
\begin{equation}
\label{eq:heat-kernel-dilation}
K(t)=E_N^{\dagger}(\bra{g}\otimes I)e^{-i\sqrt{t}\,Q\otimes B_A}
(\ket{g}\otimes I)E_N.
\end{equation}
The unitary evolution has a $\sqrt t$ scale and system-generator norm
$\norm{B_A}=\norm{A}$, while $\norm{\mathcal{G}}=\norm{A}^2$.

\subsubsection{Observable and matrix-element estimation}
\label{sec:observable-estimation}

Rall's algorithm~\cite[Lemma~5]{rall2020quantum} estimates
$\operatorname{Tr}(\varrho O)$ for a Hermitian operator with an exact
$(\alpha_O,a_O,0)$-block-encoding and a preparation circuit for a
purification of $\varrho$. With $\bar O=(I+O/\alpha_O)/2$, the construction
estimates the amplitude
\begin{equation}
\label{eq:rall-amplitude}
s=\operatorname{Tr}(\varrho\bar O)
=\frac{1+\operatorname{Tr}(\varrho O)/\alpha_O}{2}
\end{equation}
and returns
\begin{equation}
\label{eq:rall-recovery}
\xi=\alpha_O(2\widehat s-1).
\end{equation}
For $0<\epsilon<\alpha_O$ and $0<\delta_{\mathrm{fail}}<1/2$, additive
error $\epsilon$ with failure probability at most $\delta_{\mathrm{fail}}$
requires gate cost
\begin{equation}
\label{eq:rall-complexity}
O\!\left((T_O+T_{\varrho})\frac{\alpha_O}{\epsilon}
\log\frac{1}{\delta_{\mathrm{fail}}}\right).
\end{equation}
Here $T_O,T_{\varrho}$ are the block-encoding and purification-preparation
gate costs, including controls, inverses, and reflections.
Operator approximation contributes additively to the error.
The following identity reduces real matrix elements to this setting.

\begin{lemma}[Real matrix-element reduction]
\label{lem:real-matrix-element-reduction}
Let $K(t)=e^{-t\mathcal G}$, where $\mathcal G$ is real symmetric
positive semidefinite and $t\geq0$. For normalized real states
$\ket v,\ket w$, define
\[
\ket{\chi_{v,w}}=\frac{\ket0\ket v+\ket1\ket w}{\sqrt2}.
\]
Then, with $\sigma_x$ the Pauli $X$ operator,
\begin{equation}
\label{eq:matrix-element-observable}
\bra v K(t)\ket w
=\bra{\chi_{v,w}}(\sigma_x\otimes K(t))\ket{\chi_{v,w}}.
\end{equation}
\end{lemma}
The identity follows immediately from real symmetry and applies equally
to $H_\beta^{-1}$. Tagged-state preparation uses the controlled query and
input oracles in \cref{ass:state-prep}.

\subsection{Heat-Semigroup Matrix-Element Estimation}
\label{sec:heat-algorithm-analysis}

We estimate the scalar targets $z(t)$ and $m_i(t)$ in \cref{prob:heat}.
These diffusion kernels capture data-graph connectivity~\cite{kondor2002diffusion}
and support selected similarities, smoothed predictions, and group averages
or contrasts. For nonzero real vectors $\bm{a},\bm{h}$,
\[
\bm{a}^TK(t)\bm{h}=\|\bm{a}\|\,\|\bm{h}\|
\langle\widehat a|K(t)|\widehat h\rangle.
\]
Taking $\bm{a}=\bm{e}_i$ gives a selected prediction, an indicator or contrast vector
gives an aggregate, and $\bm{a}=\bm{e}_i$, $\bm{h}=\bm{e}_j$ gives $K_{ij}(t)$.

\subsubsection{Algorithm}
\label{sec:heat-algorithm}

\Cref{alg:heat-matrix-element} combines Gaussian compression,
the matrix-element reduction, and Rall's estimator. The finite Gaussian
LCU construction~\cite{kharazi2026sublinear} gives a Hermitian contraction
$\widetilde K(t)$ with an exact $(1,a_K,0)$-block-encoding $U_K$ satisfying
\begin{equation}
\label{eq:heat-block-accuracy}
\norm{\widetilde K(t)-K(t)}\leq\eta.
\end{equation}
The construction and cost under our access model are given in
\cref{ass:gaussian-heat-cost}.

\begin{algorithm}[H]
\caption{Heat-semigroup matrix-element estimation by Gaussian dilation}
\label{alg:heat-matrix-element}
\begin{algorithmic}[1]
\Require $t\geq0$, $0<\epsilon<1$, $0<\delta_{\mathrm{fail}}<1/2$,
and matrix and state-preparation oracles from
\cref{ass:matrix-access,ass:state-prep}.
\Ensure An estimate of $z(t)=\bra{u}K(t)\ket{v}$ with additive error
$\epsilon$ and failure probability at most $\delta_{\mathrm{fail}}$.
\State Construct $U_K$ with $\eta=\epsilon/4$ and prepare
$\ket{\chi_{u,v}}$ from \cref{lem:real-matrix-element-reduction}.
\State Estimate $\sigma_x\otimes\widetilde K(t)$ in this state using
Rall's procedure with accuracy $\epsilon/2$ and failure probability
$\delta_{\mathrm{fail}}$.
\State \Return the resulting estimate $\widehat z(t)$.
\end{algorithmic}
\end{algorithm}

\paragraph{Joint estimation of multiple queries.}
For the targets $m_i(t)=\bra{u_i}K(t)\ket{v}$, we place the shared
computation in the state
\begin{equation}
\label{eq:heat-common-state}
\begin{aligned}
\ket{\zeta_0}&:=\ket{0}^{\otimes a_K}\ket{0}_{\mathrm{sys}},
\qquad
\ket{\zeta_v}:=\ket{0}^{\otimes a_K}\ket{v},\\
\ket{\Psi_t}&:=
\frac{\ket{0}\ket{\zeta_0}
+\ket{1}U_K\ket{\zeta_v}}{\sqrt2}.
\end{aligned}
\end{equation}
Its preparation uses a tag Hadamard and controlled $U_v,U_K$.
The query dependence enters through the Hermitian unitary observables
\begin{equation}
\label{eq:heat-query-observables}
\begin{aligned}
W_i&:=U_{u_i}^\dagger,
\qquad V_i:=I_{\mathrm{anc}}\otimes W_i,\\
O_i&:=\ket0\!\bra1\otimes V_i
+\ket1\!\bra0\otimes V_i^\dagger,
\qquad i=1,\ldots,J,
\end{aligned}
\end{equation}
where $I_{\mathrm{anc}}$ acts on the heat-block ancillas. They satisfy
\begin{equation}
\label{eq:heat-common-expectation}
\bra{\Psi_t}O_i\ket{\Psi_t}
=\operatorname{Re}\bra{\zeta_0}V_iU_K\ket{\zeta_v}
=\operatorname{Re}\bra{u_i}\widetilde K(t)\ket{v},
\end{equation}
with bias at most $\eta$. Their controlled evolutions follow from
\begin{equation}
\label{eq:heat-observable-evolution}
\begin{aligned}
\mathcal D_i&:=\ket0\!\bra0\otimes V_i
+\ket1\!\bra1\otimes I,\\
e^{-ixO_i}
&=\mathcal D_i\bigl(R_x(2x)\otimes I\bigr)\mathcal D_i^\dagger,
\qquad R_x(2x):=e^{-ix\sigma_x}.
\end{aligned}
\end{equation}
Each evolution uses controlled $U_{u_i}$ and its inverse, together
with a tag rotation, without calling $U_v$ or $U_K$.
We apply the joint estimator of~\cite[Theorem~1]{huggins2022multiple}
with $\eta=\epsilon/4$, estimation accuracy $\nu=\epsilon/2$, and
$O(\log(1/\delta_{\mathrm{fail}}))$ independent repetitions followed
by coordinatewise medians.

\subsubsection{Error and Complexity Analysis}
\label{sec:heat-analysis}

The following lemma gives the finite Gaussian implementation under
\cref{ass:matrix-access}.

\begin{lemma}[Efficient Gaussian heat-block implementation]
\label{ass:gaussian-heat-cost}
Suppose the matrix-access model in \cref{ass:matrix-access} holds.
For every $t\geq0$ and $0<\eta<1/4$, an explicit finite Gaussian
construction supplies an exact $(1,a_K,0)$-block-encoding $U_K$
of a Hermitian contraction $\widetilde K(t)$ satisfying
\begin{equation}
\norm{\widetilde K(t)-K(t)}\leq\eta,
\qquad K(t)=e^{-t\mathcal G}.
\end{equation}
It uses $\widetilde O(1+\alpha_A\sqrt t)$ matrix-oracle calls, including inverses
and controls. Each application of $U_K$, its inverse, or a required controlled version costs
\begin{equation}
\label{eq:gaussian-heat-cost}
T_K(t,\eta)
=
\widetilde O\!\left((1+\alpha_A\sqrt t)T_A\right).
\end{equation}

Here $T_A$ includes signal-subspace reflections and embedding controls.
If it counts only the matrix-oracle circuit, replace it in
\eqref{eq:gaussian-heat-cost} by $T_A+a_A+\log(N+M)$.
\end{lemma}

The implementation follows Gaussian-LCHS~\cite{kharazi2026sublinear},
applied to the factored-generator dilation of~\cite[Section~5.3]{wu2026quantum}.
\Cref{app:gaussian-heat-cost-proof} summarizes the preparation and binary
time-selection costs under our access model. Arithmetic, simulation,
and synthesis are included. Scalar parameters are supplied in finite
binary form, and classical preprocessing is charged separately.
Here $\widetilde O$ suppresses polylogarithmic factors in $1/\eta$,
$N+M$, $a_A+2$, and $2+\alpha_A\sqrt t$.

\begin{theorem}[Matrix-element estimation by Gaussian dilation]
\label{thm:single-heat-estimation}
Let $\mathcal{G}=A^TA\succeq0$ with $A\in\mathbb{R}^{M\times N}$, and suppose
\cref{ass:matrix-access,ass:state-prep} hold.
For normalized real states $\ket{u},\ket{v}\in\mathbb{R}^N$, every $t\geq0$,
$0<\epsilon<1$, and $0<\delta_{\mathrm{fail}}<1/2$,
\cref{alg:heat-matrix-element} outputs an estimate $\widehat z(t)\in\mathbb{R}$ such that
\begin{equation}
\label{eq:single-heat-guarantee}
\Pr\!\left[\left|\widehat z(t)-\bra{u}K(t)\ket{v}\right|\leq\epsilon\right]
\geq1-\delta_{\mathrm{fail}}.
\end{equation}
Its gate complexity is
\begin{equation}
\label{eq:single-heat-complexity}
\widetilde O\!\left(
\frac{T_u+T_v+(1+\alpha_A\sqrt{t})T_A}{\epsilon}
\log\frac{1}{\delta_{\mathrm{fail}}}\right).
\end{equation}
The soft-$O$ notation includes polylogarithmic factors in $1/\epsilon$ for heat-block
accuracy $\eta=\epsilon/4$ and circuit implementation.
\end{theorem}
The result follows by applying Rall's estimator
\cite[Lemma~5]{rall2020quantum} to \eqref{eq:matrix-element-observable}.
\Cref{app:proof-single-heat-estimation} records the error allocation
and resulting cost.

\begin{theorem}[Joint estimation of heat-semigroup matrix elements]
\label{thm:joint-heat-estimation}
\label{alg:joint-heat-matrix-elements}
Let $\mathcal{G}=A^TA\succeq0$ with $A\in\mathbb{R}^{M\times N}$, and suppose
\cref{ass:matrix-access,ass:state-prep} hold.
Let $\ket{v},\ket{u_1},\ldots,\ket{u_J}\in\mathbb{R}^N$ be normalized, with $J\geq1$.
For every $t\geq0$, $0<\epsilon<1$, and $0<\delta_{\mathrm{fail}}<1/2$,
there is an algorithm that outputs real estimates
$\widehat m_1(t),\ldots,\widehat m_J(t)$ satisfying
\begin{equation}
\label{eq:joint-heat-guarantee}
\Pr\!\left[\max_{1\leq i\leq J}
\left|\widehat m_i(t)-\bra{u_i}K(t)\ket{v}\right|\leq\epsilon\right]
\geq1-\delta_{\mathrm{fail}}.
\end{equation}
The number of calls to $U_K$ and its inverse, including their controlled versions, and the
gate cost of the common-state preparations are, respectively,
\begin{equation}
\label{eq:joint-heat-common-cost}
\begin{aligned}
n_K^{(J)}&=\widetilde O\!\left(\frac{\sqrt{J}}{\epsilon}
\log\frac{1}{\delta_{\mathrm{fail}}}\right),\\
T_{\mathrm{common}}^{(J)}&=\widetilde O\!\left(\frac{\sqrt{J}}{\epsilon}
\bigl[T_v+(1+\alpha_A\sqrt{t})T_A\bigr]
\log\frac{1}{\delta_{\mathrm{fail}}}\right).
\end{aligned}
\end{equation}
Here $\widetilde O$ also suppresses logarithmic factors in $J$ and $1/\epsilon$.
The factor $\log(1/\delta_{\mathrm{fail}})$ accounts for confidence amplification,
and $T_{\mathrm{common}}^{(J)}$ includes common-state preparation and its inverse.
\end{theorem}
The result follows from the joint estimator of
\cite[Theorem~1]{huggins2022multiple} applied to
\cref{eq:heat-common-state,eq:heat-query-observables}.
\Cref{app:proof-joint-heat-estimation} verifies the reduction and
accounts for observable-processing costs.

\subsection{Regularized Inverse Scalar Estimation}
\label{sec:inverse-algorithm-analysis}

We now turn to estimating $r_\beta(\bm{y},\bm{b})=\bra{y}(\mathcal{G}+\beta I)^{-1}\ket{b}$
from \cref{prob:inverse}, with $\beta>0$. In particular, we construct the normalized resolvent
\begin{equation}
\label{eq:normalized-resolvent}
R_\beta:=\beta H_\beta^{-1}=\beta(\mathcal{G}+\beta I)^{-1}, \qquad
0\preceq R_\beta\preceq I, \qquad
r_\beta(\bm{y},\bm{b})=\frac{1}{\beta}\bra{y}R_\beta\ket{b}.
\end{equation}
The block normalization is one, and the final estimate is divided by $\beta$.

\subsubsection{The quantum algorithm}
\label{sec:inverse-algorithm}

The Laplace-transform framework~\cite{an2026laplace} gives
\begin{equation}
\label{eq:inverse-laplace-transform}
R_\beta=\int_0^\infty\beta e^{-\beta t}K(t)\,dt.
\end{equation}
Combining this identity with the Gaussian dilation
\eqref{eq:heat-kernel-dilation} yields the single-integral representation
\begin{equation}
\label{eq:square-root-resolvent}
R_\beta
=E_N^\dagger\!\left(
\int_{\mathbb R}p_\beta(k)e^{-ikB_A}\,dk
\right)E_N,
\qquad
p_\beta(k):=\frac{\sqrt\beta}{2}e^{-\sqrt\beta|k|}.
\end{equation}
The normalized Laplace density has exponentially decaying tails, so
\begin{equation}
\label{eq:resolvent-tail}
R_{\beta,\Lambda}
:=E_N^\dagger\!\left(
\int_{-\Lambda}^{\Lambda}p_\beta(k)e^{-ikB_A}\,dk
\right)E_N,
\qquad
\norm{R_\beta-R_{\beta,\Lambda}}
\leq e^{-\sqrt\beta\Lambda}.
\end{equation}
For $0<\eta<1/4$, set
\begin{equation}
\label{eq:resolvent-cutoff}
\Lambda_{\beta,\eta}:=\frac{1}{\sqrt\beta}\ln\frac{8}{\eta},
\qquad
\bar p_{\beta,\Lambda}(k)
:=\frac{p_\beta(k)}{1-e^{-\sqrt\beta\Lambda}},
\quad |k|\leq\Lambda.
\end{equation}
Following the quadrature and LCU framework of
\cite[Sections~3.2--3.5 and Appendix~B]{an2026laplace}, a normalized
nonnegative quadrature of this density gives an exact
$(1,a_R,0)$-block-encoding $U_R$ of a Hermitian contraction
$\widetilde R_\beta$ satisfying
\begin{equation}
\label{eq:inverse-block-accuracy}
\norm{\widetilde R_\beta-R_\beta}\leq\eta.
\end{equation}
The resulting cost is stated in \cref{thm:inverse-block-encoding}.
\Cref{app:resolvent-representation,app:explicit-laplace-quadrature}
give the scalar integral identity and geometric-weight specialization.

Using the tagged state from \cref{sec:observable-estimation},
\begin{equation}
\label{eq:inverse-tagged-expectation}
\bra{\chi_{y,b}}
(\sigma_x\otimes\widetilde R_\beta)
\ket{\chi_{y,b}}
=\operatorname{Re}\bra{y}\widetilde R_\beta\ket{b}.
\end{equation}
\Cref{alg:inverse-scalar} applies Rall's estimator~\cite{rall2020quantum}
with tolerances scaled by $\beta$ and divides the result by $\beta$.
If $\beta\epsilon\geq1$, the bound $|r_\beta(\bm{y},\bm{b})|\leq1/\beta$
makes zero a valid output.

\begin{algorithm}[H]
\caption{Regularized inverse scalar estimation by square-root Lap-LCHS}
\label{alg:inverse-scalar}
\begin{algorithmic}[1]
\Require $\beta>0$, $\epsilon>0$, $0<\delta_{\mathrm{fail}}<1/2$,
and oracles $U_B,U_y,U_b$ from \cref{ass:matrix-access,ass:state-prep}.
\Ensure An estimate of $r_\beta(\bm{y},\bm{b})$ with additive error $\epsilon$
and failure probability at most $\delta_{\mathrm{fail}}$.
\If{$\beta\epsilon\geq1$}
\State \Return $0$.
\EndIf
\State Construct $U_R$ with $\eta=\beta\epsilon/4$ and prepare
$\ket{\chi_{y,b}}$.
\State Estimate $\sigma_x\otimes\widetilde R_\beta$ in this state
with accuracy $\beta\epsilon/2$ and failure probability
$\delta_{\mathrm{fail}}$, obtaining $\widehat o_\beta$.
\State \Return $\widehat o_\beta/\beta$.
\end{algorithmic}
\end{algorithm}

As in the heat problem, joint estimation uses a common state and
query-dependent observables~\cite{huggins2022multiple}. In
\cref{eq:heat-common-state,eq:heat-query-observables,eq:heat-common-expectation},
replace $U_K,U_v,U_{u_i}$ by $U_R,U_b,U_{y_i}$, use
$\eta=\beta\epsilon/4$ and $\nu=\beta\epsilon/2$, and divide the estimates
by $\beta$.

\subsubsection{Error and Complexity Analysis}
\label{sec:inverse-analysis}

Hamiltonian simulation~\cite[full version, Theorem~58]{gilyen2019quantum} implements
$e^{-i\tau B_A}$ to error $\zeta$ with
\begin{equation}
\label{ass:inverse-hamiltonian-simulation}
\widetilde O(1+\alpha_A|\tau|)\quad\text{matrix-oracle calls},
\qquad
\widetilde O\!\left((1+\alpha_A|\tau|)T_A\right)\quad\text{gates}.
\end{equation}
The suppressed factors include precision, register sizes, and controls.
The cutoff in \eqref{eq:resolvent-cutoff} and the coherent quadrature in
\cref{lem:appendix-explicit-laplace-quadrature} give the following bound.

\begin{theorem}[Square-root Lap-LCHS resolvent block encoding]
\label{thm:inverse-block-encoding}
Suppose \cref{ass:matrix-access} holds. For every $\beta>0$ and $0<\eta<1/4$, the construction in
\cref{sec:inverse-algorithm} produces an exact $(1,a_R,0)$-block-encoding $U_R$ of a
Hermitian contraction $\widetilde R_\beta$ satisfying
$\norm{\widetilde R_\beta-R_\beta}\leq\eta$. Its number of queries to $U_B,U_B^\dagger$
and its gate cost, including controlled and inverse versions, are
\begin{equation}
\label{eq:inverse-block-cost}
n_{B,R}(\beta,\eta)=\widetilde O\!\left(1+\frac{\alpha_A}{\sqrt{\beta}}\right),
\qquad
T_R(\beta,\eta)=\widetilde O\!\left(\left(1+\frac{\alpha_A}{\sqrt{\beta}}\right)T_A\right).
\end{equation}
Suppressed factors are polylogarithmic in $1/\eta$, quadrature parameters, and register sizes.
\end{theorem}
This specializes the cited Lap-LCHS implementation to
\eqref{eq:square-root-resolvent}.
\Cref{app:proof-inverse-block-encoding} combines the quadrature,
preparation, and simulation errors to obtain \eqref{eq:inverse-block-cost}.

\begin{theorem}[Regularized inverse scalar estimation]
\label{thm:inverse-scalar-estimation}
Suppose \cref{ass:matrix-access,ass:state-prep} hold. For normalized real states $\ket{y},\ket{b}\in\mathbb{R}^N$, $\beta>0$,
$\epsilon>0$, and $0<\delta_{\mathrm{fail}}<1/2$, \cref{alg:inverse-scalar} outputs
$\widehat r_\beta(\bm{y},\bm{b})\in\mathbb{R}$ satisfying
\begin{equation}
\label{eq:inverse-scalar-guarantee}
\Pr\!\left[\left|\widehat r_\beta(\bm{y},\bm{b})-\bra{y}H_\beta^{-1}\ket{b}\right|
\leq\epsilon\right]\geq1-\delta_{\mathrm{fail}}.
\end{equation}
In the nontrivial regime $0<\beta\epsilon<1$, its gate complexity and number of calls
to the matrix oracle $U_B$ and its inverse are, respectively,
\begin{equation}
\label{eq:inverse-scalar-cost}
\begin{aligned}
T_{\mathrm{scalar},R}&=\widetilde O\!\left(
\frac{T_y+T_b+(1+\alpha_A/\sqrt{\beta})T_A}{\beta\epsilon}
\log\frac{1}{\delta_{\mathrm{fail}}}\right),\\
n_{B,\mathrm{scalar},R}&=\widetilde O\!\left(
\frac{1+\alpha_A/\sqrt{\beta}}{\beta\epsilon}
\log\frac{1}{\delta_{\mathrm{fail}}}\right).
\end{aligned}
\end{equation}
The soft-$O$ notation includes polylogarithmic factors in $1/(\beta\epsilon)$
from $\eta=\beta\epsilon/4$. For $\beta\epsilon\geq1$, zero satisfies the guarantee
without oracle calls.
\end{theorem}
The result follows from Rall's estimator
\cite[Lemma~5]{rall2020quantum} applied to
\eqref{eq:inverse-tagged-expectation}, followed by division by $\beta$.
\Cref{app:proof-inverse-scalar-estimation} records the rescaled
error and cost bounds.

\begin{theorem}[Joint regularized inverse scalar estimation]
\label{thm:joint-inverse-estimation}
\label{alg:joint-inverse-scalars}
Suppose \cref{ass:matrix-access,ass:state-prep}
hold. Let $\ket b\in\mathbb R^N$ be normalized, with normalized real queries
$\ket{y_i}$ for $i=1,\ldots,J$, where $J\geq1$.
For every $\beta>0$, $\epsilon>0$, and
$0<\delta_{\mathrm{fail}}<1/2$,
the common-state construction above outputs real estimates satisfying
\begin{equation}
\label{eq:joint-inverse-guarantee}
\Pr\!\left[\max_{1\leq i\leq J}
\left|\widehat r_{\beta,i}-r_\beta(\bm{y}_i,\bm{b})\right|\leq\epsilon\right]
\geq1-\delta_{\mathrm{fail}}.
\end{equation}
For $0<\beta\epsilon<1$, the number of calls to $U_R$ and its inverse, including controlled
versions, and the gate cost of the common-state preparations are
\begin{equation}
\label{eq:joint-inverse-common-cost}
\begin{aligned}
n_R^{(J)}&=\widetilde O\!\left(\frac{\sqrt{J}}{\beta\epsilon}
\log\frac{1}{\delta_{\mathrm{fail}}}\right),\\
T_{\mathrm{common},R}^{(J)}&=\widetilde O\!\left(\frac{\sqrt{J}}{\beta\epsilon}
\bigl[T_b+(1+\alpha_A/\sqrt{\beta})T_A\bigr]
\log\frac{1}{\delta_{\mathrm{fail}}}\right).
\end{aligned}
\end{equation}
The suppressed logarithmic factors additionally depend on $J$ and $1/(\beta\epsilon)$.
For $\beta\epsilon\geq1$, the all-zero output satisfies the same guarantee without oracle calls.
\end{theorem}
The result applies the joint estimator of
\cite[Theorem~1]{huggins2022multiple} to the common resolvent state.
The rescaling and cost calculation are given in
\cref{app:proof-joint-inverse-estimation}.

With $S_y:=\sum_{i=1}^J(T_{y_i}+T_{\mathrm{rot}})$, where $T_{\mathrm{rot}}$
includes controlled rotations and synthesis, and $\eta=\beta\epsilon/4$,
the total gate cost for $0<\beta\epsilon<1$ is
\begin{equation}
\label{eq:joint-inverse-total-cost}
T_{\mathrm{joint},R}^{(J)}
=\widetilde O\!\left(\frac{\sqrt J}{\beta\epsilon}
\bigl[T_b+T_R(\beta,\eta)+S_y\bigr]
\log\frac1{\delta_{\mathrm{fail}}}\right).
\end{equation}
Classical output processing is additional. Independent readout follows
\eqref{eq:heat-independent-readout-costs} with
$C_t\mapsto T_b+T_R(\beta,\eta)$, $S_u\mapsto S_y$, and
$\epsilon\mapsto\beta\epsilon$. 

\subsection{Spectral Galerkin State Preparation}
\label{sec:galerkin-algorithm-analysis}

For the energy minimizer $\bm{u}_\delta=L^+P_\delta^+\bm{f}\in\mathcal{E}_\delta$
in \cref{prob:galerkin}, with $L=\mathcal{G}=A^TA$,
we construct a bounded projected inverse by QSVT~\cite{gilyen2019quantum}.
Applied to the source state in \cref{ass:source-oracle}, it prepares the
normalized solution by fixed-point amplitude amplification~\cite{yoder2014fixed}.

The retained projector commutes with $L$, so a scalar spectral transformation
gives the Galerkin solution. A separate diffusion-map basis $U_r$ for an
operator $L_{\mathrm{PDE}}$ instead gives
\[
\bm{u}_r=U_r(U_r^TL_{\mathrm{PDE}}U_r)^{-1}U_r^T\bm{f},
\]
when the reduced matrix is positive definite. This broader setting requires
access to the compressed operator~\cite{yan2023spectral}. Here we use the
commuting spectral restriction in \cref{sec:setup-galerkin}.

\paragraph{Learning interpretation.}
Low-frequency graph regression minimizes
\[
\mathcal J_\beta(\bm{z})=\frac12\bm{z}^T\mathcal G_0\bm{z}+
\frac\beta2\|\bm{z}-\bm{\ell}\|^2,
\qquad \bm{z}\in\operatorname{Ran}\mathbf1_{[0,\theta]}(\mathcal G_0),
\]
combining geometric smoothness and spectral restriction
\cite{zhou2003consistency,belkin2006manifold}.
With $\mathcal G_0=A_0^TA_0$, this has the present form for
$L=\mathcal G_0+\beta I$, $A=(A_0^T,\sqrt\beta I)^T$,
$\bm{f}=\beta\bm{\ell}$, and cutoff $\delta=\theta+\beta$.
Poisson learning similarly uses labels as sources and sinks in a graph
Poisson equation~\cite{calder2020poisson}, with spectral restriction
providing an additional approximation.

\subsubsection{Algorithm}
\label{sec:galerkin-algorithm}
\label{thm:galerkin-projected-inverse}

Let $\gamma>0$ bound the positive spectrum of $L$ from below, and let
$\delta_-<\delta<\delta_+$ define the empty cutoff interval in
\cref{ass:galerkin-spectrum}. When all positive modes are retained,
omit the upper cutoff filter and $\delta_+$. Define
\begin{equation}
\label{eq:galerkin-projected-inverse}
Q_\delta:=\frac{\gamma}{4}L^+P_\delta^+
=\frac{\gamma}{4}B_\delta^+, \qquad
0\preceq Q_\delta\preceq\frac14 I.
\end{equation}
This operator inverts the retained modes and removes the others. On the
normalized source,
\begin{equation}
\label{eq:galerkin-good-vector}
Q_\delta\ket{\hat f}
=\frac{\gamma}{4F}\sum_{i=1}^N(\bm{u}_\delta)_i\ket{i}, \qquad
s_\delta:=\norm{Q_\delta\ket{\hat f}}
=\frac{\gamma\norm{\bm{u}_\delta}}{4F}.
\end{equation}
The normalized successful branch is the desired solution state.

Choose a real odd polynomial $p$, bounded by one on $[-1,1]$, satisfying
\begin{equation}
\label{eq:galerkin-reciprocal-polynomial}
\begin{aligned}
\left|p(\sigma/\alpha_A)-\frac{\sqrt{\gamma}}{2\sigma}\right|
&\leq\eta_p, &&\sqrt{\gamma}\leq\sigma\leq\sqrt{\delta_-},\\
|p(\sigma/\alpha_A)|&\leq\eta_p,
&&\sqrt{\delta_+}\leq\sigma\leq\alpha_A.
\end{aligned}
\end{equation}
Standard reciprocal and cutoff approximations give such a bounded
polynomial~\cite{gilyen2019quantum}. Its square $q=p^2$ is even, vanishes
at zero, and implements the projected inverse through
\eqref{eq:dilation-square}:

\begin{equation}
\label{eq:galerkin-even-transform}
Q_{\delta,p}:=E_N^{\dagger}q(B_A/\alpha_A)E_N
=\sum_{j=1}^N p(\sqrt{\lambda_j}/\alpha_A)^2 \bm{q}_j\bm{q}_j^T,
\qquad \norm{Q_{\delta,p}-Q_\delta}\leq\eta_p+\eta_p^2.
\end{equation}
Here $L\bm{q}_j=\lambda_j\bm{q}_j$, as in \cref{sec:setup-galerkin}.
Squaring supplies the $1/\lambda_j$ factor without eigenvector preparation.

For $0<\eta_Q<1$, take $\eta_p\leq\eta_Q/4$ and implementation error at most
$\eta_Q/2$. A constant number of signal ancillas and Hermitian symmetrization,
when needed, give a unitary $U_Q$ with
\begin{equation}
\label{eq:galerkin-implemented-block}
(\bra{0}^{\otimes a_Q}\otimes I)U_Q(\ket{0}^{\otimes a_Q}\otimes I)
=\widetilde Q_\delta, \qquad
\widetilde Q_\delta=\widetilde Q_\delta^\dagger, \qquad
\norm{\widetilde Q_\delta}\leq1, \quad
\norm{\widetilde Q_\delta-Q_\delta}\leq\eta_Q.
\end{equation}
Here $a_Q$ includes the embedding and signal registers.

For $W_Q=U_Q(I\otimes U_f)$, the all-zero block-ancilla outcome gives
$\widetilde Q_\delta\ket{\hat f}$. Given $s_0\leq s_\delta$, fixed-point
amplification uses $W_Q$, $W_Q^\dagger$, and phase rotations about the initial
and successful subspaces~\cite{yoder2014fixed}. We then measure the success flag.
\begin{definition}[Distance up to a global phase]
\label{def:phase-distance}
For normalized quantum states $\ket{\psi}$ and $\ket{\phi}$ in the same
Hilbert space, define
\begin{equation}
\label{eq:galerkin-state-distance}
d_{\mathrm{ph}}(\ket{\psi},\ket{\phi})
:=\min_{\vartheta\in\mathbb{R}}
\norm{\ket{\psi}-e^{i\vartheta}\ket{\phi}}_2.
\end{equation}

\end{definition}

\begin{algorithm}[H]
\caption{Spectral Galerkin solution-state preparation}
\label{alg:galerkin-state}
\begin{algorithmic}[1]
\Require Oracles $U_B,U_f$, spectral data $\gamma,\delta_-,\delta,\delta_+$,
known $0<s_0\leq s_\delta$, $0<\epsilon_{\mathrm{st}}<1$, and
$0<\delta_{\mathrm{fail}}<1/2$.
\Ensure A success flag with probability at least $1-\delta_{\mathrm{fail}}$ and,
conditional on success, a state $\ket{\psi_\delta}$ satisfying
$d_{\mathrm{ph}}(\ket{\psi_\delta},\ket{\hat u_\delta})\leq\epsilon_{\mathrm{st}}$.
\State Set $\eta_Q\gets\epsilon_{\mathrm{st}}s_0/8$.
\State Construct $U_Q$ using \eqref{eq:galerkin-reciprocal-polynomial}--\eqref{eq:galerkin-implemented-block}
with tolerance $\eta_Q$.
\State Form $W_Q\gets U_Q(I\otimes U_f)$ and mark the all-zero block-ancilla subspace.
\State Apply fixed-point amplification with amplitude lower bound $s_0/2$ and
ideal failure probability at most $\delta_{\mathrm{fail}}/2$, using
success-probability lower bound $(s_0/2)^2$.
\State Compile the amplified circuit to operator-norm error at most
$\min\{\epsilon_{\mathrm{st}}/16,\delta_{\mathrm{fail}}/8\}$ relative to that circuit.
\State Measure the success flag. Return the system register on success, or report failure.
\end{algorithmic}
\end{algorithm}

\subsubsection{Error and Complexity Analysis}
\label{sec:galerkin-analysis}

We use \cref{ass:matrix-access,ass:source-oracle} with the following promises.
The cutoff is $\delta$ and the failure tolerance is $\delta_{\mathrm{fail}}$.

\begin{assumption}[Spectral Galerkin access and parameters]
\label{ass:galerkin-spectrum}
\label{ass:galerkin-overlap}
\label{ass:galerkin-controls}
\emph{Spectrum.} Known parameters satisfy
\begin{equation}
\label{eq:galerkin-spectral-promise}
0<\gamma\leq\delta_-<\delta<\delta_+\leq\alpha_A^2, \qquad
\operatorname{spec}(L)\subseteq
\{0\}\cup[\gamma,\delta_-]\cup[\delta_+,\alpha_A^2].
\end{equation}
Set $\Delta_\sigma:=\sqrt{\delta_+}-\sqrt{\delta_-}>0$.
If all positive modes are retained, use
$\operatorname{spec}(L)\subseteq\{0\}\cup[\gamma,\delta_-]$ with
$0<\gamma\leq\delta_-\leq\min\{\delta,\alpha_A^2\}$, and omit
$\delta_+$, the upper cutoff filter, and all $\Delta_\sigma^{-1}$ terms.

\emph{Implementation.} Signal rotations and embeddings cost $T_{\mathrm{sig}}$
per polynomial step. Initial-state and successful-subspace rotations cost
$T_{\mathrm{ref}}$ per amplification or estimation step. Both costs include
precision and controls. Classical computation of polynomial and amplification phases
is charged separately.

\emph{State preparation only.} Assume $P_\delta^+\bm{f}\neq0$ and a known
$0<s_0\leq s_\delta$. A known bound
$0<r_0\leq r_\delta:=\norm{P_\delta^+\ket{\hat f}}$ also suffices, since
\begin{equation}
\label{eq:galerkin-overlap-bound}
\frac{\gamma r_\delta}{4\delta_-}\leq s_\delta\leq\frac{r_\delta}{4},
\qquad s_0=\frac{\gamma r_0}{4\delta_-}
\end{equation}
is valid. The amplified circuit has operator-norm synthesis error at most
$\min\{\epsilon_{\mathrm{st}}/16,\delta_{\mathrm{fail}}/8\}$.

\end{assumption}

\paragraph{Projected-inverse construction and cost.}
The reciprocal and cutoff polynomials, followed by real-polynomial QSVT,
give \eqref{eq:galerkin-implemented-block} for every $0<\eta_Q<1$
\cite[full version, Corollaries~18 and~67, Lemma~29]{gilyen2019quantum}.
This specializes the threshold-pseudoinverse construction in
\cite[full version, Corollary~42]{gilyen2019quantum} to the retained low positive band,
with the squared polynomial implementing $L^+$.
The degree and gate cost are
\begin{equation}
\label{eq:galerkin-qsvt-degree}
d_Q(\eta_Q)=\widetilde O\!\left(
\frac{\alpha_A}{\sqrt{\gamma}}+\frac{\alpha_A}{\Delta_\sigma}\right),
\end{equation}
\begin{equation}
\label{eq:galerkin-block-cost}
T_Q(\eta_Q)=O\!\left(d_Q(\eta_Q)(T_A+T_{\mathrm{sig}})\right).
\end{equation}
There are $O(d_Q(\eta_Q))$ matrix-oracle calls, including inverses and controls.
The soft-$O$ notation hides logarithmic factors in $1/\eta_Q$ and the inverse
normalized spectral gaps. The error allocation is verified in
\cref{app:proof-galerkin-projected-inverse}.

\begin{theorem}[Spectral Galerkin solution-state preparation]
\label{thm:galerkin-state}
Assume \cref{ass:matrix-access,ass:source-oracle,ass:galerkin-spectrum},
including the state-preparation conditions.
For every $0<\epsilon_{\mathrm{st}}<1$ and $0<\delta_{\mathrm{fail}}<1/2$,
\cref{alg:galerkin-state} returns a success flag with probability at least
$1-\delta_{\mathrm{fail}}$. Conditional on success, its output satisfies
\begin{equation}
\label{eq:galerkin-state-guarantee}
d_{\mathrm{ph}}(\ket{\psi_\delta},\ket{\hat u_\delta})\leq\epsilon_{\mathrm{st}}.
\end{equation}
Here $d_{\mathrm{ph}}$ is the distance up to a global phase from \cref{def:phase-distance}.
With $\eta_Q=\epsilon_{\mathrm{st}}s_0/8$, its matrix-oracle and source-oracle
query counts, including inverses and controls, are
\begin{equation}
\label{eq:galerkin-state-queries}
n_{B,\mathrm{st}}=O\!\left(\frac{d_Q(\eta_Q)}{s_0}
\log\frac{2}{\delta_{\mathrm{fail}}}\right), \qquad
n_{f,\mathrm{st}}=O\!\left(\frac{1}{s_0}
\log\frac{2}{\delta_{\mathrm{fail}}}\right).
\end{equation}
Its gate complexity is
\begin{equation}
\label{eq:galerkin-state-cost}
T_{\mathrm{st}}=O\!\left(
\frac{d_Q(\eta_Q)(T_A+T_{\mathrm{sig}})+T_f+T_{\mathrm{ref}}}{s_0}
\log\frac{2}{\delta_{\mathrm{fail}}}\right).
\end{equation}
Under the alternative overlap promise $r_\delta\geq r_0>0$, one may substitute
$s_0=\gamma r_0/(4\delta_-)$ in these bounds.
\end{theorem}
The result combines the QSVT construction above with fixed-point
amplitude amplification~\cite{yoder2014fixed}.
\Cref{app:proof-galerkin-state} verifies the conditional state error,
success probability, and oracle counts.

\subsection{Analytical resource comparison with classical solvers}
\label{sec:numerical-heat-comparison}

To contextualize our complexity guarantees, we compare the single-overlap
heat estimator of \cref{thm:single-heat-estimation} against representative
classical algorithms for evaluating $z(t)=\bra{u}e^{-t\mathcal G}\ket{v}$.
Standard explicit-vector methods approximate the matrix-exponential action
$e^{-t\mathcal G}\bm{v}$ and take an inner product with $\bm{u}$. Letting $C_{\mathcal G}$
denote the arithmetic cost of one matrix--vector product with $\mathcal G$ and
$k$ denote the method-dependent polynomial or Krylov step count required for
additive accuracy $\epsilon$, \cref{tab:heat-dimension-comparison} summarizes
the resulting scaling. The cited classical methods and accuracy
conventions are summarized in
\cref{app:heat-classical-comparison}.

\begin{table}[htbp]
\centering
\small
\caption{Analytical resource comparison for estimating one heat matrix
element at additive accuracy $\epsilon$. Classical rows operate on explicit
$N$-dimensional vectors with method-dependent iteration count $k$. The final
column fixes accuracy, confidence, and spectral scales, assuming
$C_{\mathcal G}=O(N)$ for generic sparse matrices. The quantum row additionally
assumes $T_A,T_u,T_v=\operatorname{polylog}(N)$, bounded $\alpha_A\sqrt t$,
and the Gaussian dilation circuit of \cref{thm:single-heat-estimation}, with
logarithmic factors in precision and confidence suppressed.}
\label{tab:heat-dimension-comparison}
\begin{tabularx}{\textwidth}{@{}>{\raggedright\arraybackslash}p{0.29\textwidth}X>{\raggedright\arraybackslash}p{0.18\textwidth}@{}}
\toprule
Method & Arithmetic or gate cost & Dependence on $N$\\
\midrule
Chebyshev recurrence~\cite{sachdeva2014faster}
& $O(k(C_{\mathcal G}+N))$ & $O(N)$\\
Lanczos~\cite{hochbruck1997krylov}
& $O(k(C_{\mathcal G}+N)+k^2)$ & $O(N)$\\
Scaled Taylor action~\cite{almohy2011action}
& $O(k(C_{\mathcal G}+N))$ & $O(N)$\\
Rational Krylov~\cite{guettel2013rational}
& Shifted solves plus $O(Nk^2+k^3)$ & Solver dependent\\
FFT for circulant $\mathcal G$
& $O(N\log N)$ & $O(N\log N)$\\
Gaussian dilation, \cref{thm:single-heat-estimation}
& $\begin{aligned}
\widetilde O\bigl(&\epsilon^{-1}[T_u+T_v\\
&{}+(1+\alpha_A\sqrt t)T_A]\bigr)
\end{aligned}$
& $\operatorname{polylog}(N)$\\
\bottomrule
\end{tabularx}
\end{table}

In \cref{tab:heat-dimension-comparison}, the Lanczos entry reflects the
three-term recurrence in exact arithmetic, whereas full reorthogonalization
incurs an additional $O(Nk^2)$ overhead. Notably, classical polynomial
methods already achieve square-root scaling in $t\norm{\mathcal G}$. The
quantum advantage lies instead in replacing $O(N)$ explicit-vector updates
with coherent block-encoding access and scalar readout. Under bounded spectral
scales and polylogarithmic oracle costs, this reduces the dimension dependence
at fixed normalized-overlap accuracy from $O(N)$ to $\operatorname{polylog}(N)$,
while state-preparation costs, block normalization $\alpha_A$, and the
$\epsilon^{-1}$ estimation scaling remain explicit.

Two caveats delineate the boundary of this asymptotic separation. First,
when an application-level observable takes the rescaled form
$Q_N=c_N\langle u|K(t)|v\rangle$, achieving absolute tolerance
$\epsilon_{\rm out}$ requires normalized-overlap accuracy
$\epsilon=\epsilon_{\rm out}/|c_N|$, introducing a factor of
$|c_N|/\epsilon_{\rm out}$ into the query complexity. Consequently, raw
nodal predictions or continuum kernel densities do not automatically inherit
polylogarithmic scaling unless their normalization factors remain bounded.
Second, \cref{tab:heat-dimension-comparison} benchmarks against generic
full-vector solvers rather than establishing an unconditional classical lower
bound: localized graph-diffusion routines~\cite{kloster2014heat} and
Poisson-length random-walk estimators (which use $O((1+t)\epsilon^{-2})$ local
steps, as summarized in \cref{app:heat-classical-comparison}) can also avoid $O(N)$
traversal for localized queries, and randomized sketching can reduce classical
Krylov orthogonalization overhead~\cite{guettel2023sketching}.

\section{Numerical Examples}
\label{sec:numerical-discussions}
We validate the discrete operator constructions and scalar readout protocols
through noiseless statevector simulations and exact estimator distributions.
Specifically, \cref{sec:numerical-circle} verifies circuit fidelity and
continuum PDE convergence on a discretized circle, while
\cref{sec:numerical-query-advantage} isolates the quadratic precision
advantage of coherent amplitude estimation under matched oracle access.
The supporting cost and error analysis is given in
\cref{app:numerical-circuits}. 

\subsection{Circuit validation and a PDE on the circle}
\label{sec:numerical-circle}

We first verify the discrete circuit constructions against explicit reference
block encodings on small instances where exact statevector simulation is
tractable. For the heat-kernel task, we use four-point Gauss--Hermite
quadrature for \eqref{eq:gaussian-integral} and a real matrix-element
Hadamard test for \eqref{eq:matrix-element-observable}.
The four-node cycle experiment uses the incidence matrix and graph Laplacian
\[
A=\begin{pmatrix}
-1&1&0&0\\
0&-1&1&0\\
0&0&-1&1\\
1&0&0&-1
\end{pmatrix},\qquad
\mathcal G=A^TA=\begin{pmatrix}
2&-1&0&-1\\
-1&2&-1&0\\
0&-1&2&-1\\
-1&0&-1&2
\end{pmatrix}.
\]
At $t=0.3$, the circuit reproduces the four-point quadrature operator
$\widetilde K(t)$ to $6.12\times10^{-16}$ in spectral norm.
With $u=(1,0,0,0)^T$ and $v=(0,1,0,0)^T$, the Hadamard test reads out
the first-row, second-column entry. Its circuit expectation,
$\bra{u}\widetilde K(0.3)\ket{v}=0.1775329905$, agrees with the exact
heat matrix element
$\bra{u}e^{-0.3\mathcal G}\ket{v}=(1-e^{-1.2})/4=0.1747014470$
up to the quadrature bias of $2.83\times10^{-3}$ (values rounded).
\Cref{app:numerical-heat-validation}
summarizes this error separation. The asymptotic implementation cost follows
from the Gaussian construction in \cref{ass:gaussian-heat-cost}.

To test the spectral Galerkin solver on a continuum manifold problem, we
consider the Poisson equation $-v_\star''=g$ on the unit circle
$\mathcal M=S^1$ ($\rho=\kappa=1$, $c=0$) with exact solution and forcing
\[
v_\star(\theta)=\sin\theta+\tfrac14\cos(2\theta),\qquad
g(\theta)=\sin\theta+\cos(2\theta).
\]
Discretizing at uniform grid points $\theta_i=2\pi i/N$ with spacing
$h=2\pi/N$ and oriented cycle incidence matrix $C$ yields
\begin{equation}
\label{eq:numerical-circle-coordinates}
\mathsf M_N=hI,\qquad A=h^{-1}C,\qquad L=A^TA,\qquad
f_i=\sqrt h\,g(\theta_i),\qquad z_{\delta,i}=(\bm{u}_\delta)_i/\sqrt h.
\end{equation}
Setting the spectral cutoff to $\delta=6$ retains both nonconstant Fourier
modes present in $g$, eliminating spectral-truncation error. For $N=16$ and
odd filter degree $d=\deg p=127$, the QSP circuit implementing $p^2$ prepares
a conditional state achieving fidelity $0.9999999368$ (and phase-aligned
distance $2.51\times10^{-4}$) relative to the discrete target $\ket{\hat u_\delta}$.
The unamplified success probability is $0.0309462617$, agreeing with the
ideal-filter value ($0.0308929595$) up to polynomial approximation error,
while the circuit-to-polynomial branch discrepancy is $3.45\times10^{-16}$.
This $\sim 3.1\%$ raw success rate corresponds to $s_\delta\approx 0.176$ and
gives a constant overlap factor in the amplification cost of
\cref{thm:galerkin-state}.

\begin{figure}[htbp]
\centering
\includegraphics[width=\textwidth]{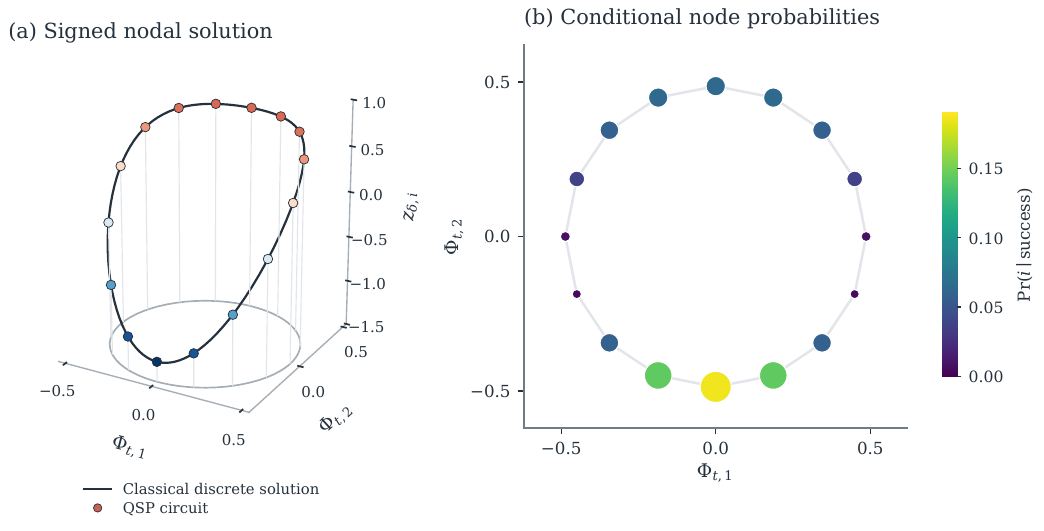}
\caption{Spectral Galerkin state on the $N=16$ circle, achieving fidelity
$0.99999994$ to the discrete target.
(a) Classical reference and simulator-reconstructed signed nodal values.
(b) Conditional node probabilities
$h|z_{\delta,i}|^2/\sum_j h|z_{\delta,j}|^2$,
indicated by color and marker area.
The signed values and diffusion-map embedding are classical diagnostics,
whereas panel~(b) displays the ideal measurement distribution conditional
on success.}
\label{fig:numerical-solution-geometry}
\end{figure}

\Cref{fig:numerical-solution-geometry} illustrates both the reconstructed
signed nodal profile and the conditional measurement distribution across the
embedded circle. Separately from circuit fidelity against $\bm{u}_\delta$, classical
grid refinement confirms second-order convergence to the continuum solution
$v_\star$, with nodal root-mean-square error decreasing from
$1.31\times10^{-2}$ at $N=16$ as $O(h^2)$. Note that because $\alpha_A=2/h=\Theta(N)$
under uniform grid refinement, this PDE sequence tests filter correctness and
discretization accuracy rather than the fixed-scale $\operatorname{polylog}(N)$
regime of \cref{tab:heat-dimension-comparison}. Circle identities are given in
\cref{app:numerical-example-identities}. 

\subsection{Quantum advantage in scalar-estimation queries}
\label{sec:numerical-query-advantage}

\begin{figure}[tbp]
\centering
\includegraphics[width=\textwidth]{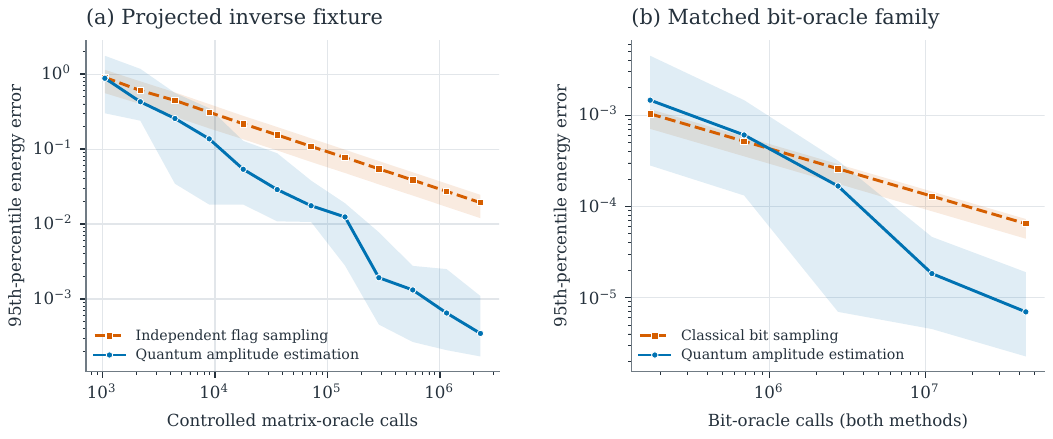}
\caption{Scalar energy estimation at matched query budgets. Curves show the
median 95th-percentile absolute energy error across instances, with
10th--90th percentile shaded bands. (a) Coherent amplitude estimation versus
independent flag sampling for sixteen complex sources in the projected-inverse
test. (b) Seven-repeat quantum amplitude estimation versus randomized
classical bit sampling for eleven means in \eqref{eq:numerical-bit-family}
at $N=2^{36}$. Distributions are exact and validated by full circuits at
$N=4,8$.}
\label{fig:numerical-query-comparison}
\end{figure}

For Galerkin energy estimation, a single odd reciprocal filter $p$ suffices.
We apply canonical amplitude estimation~\cite{brassard2002quantum} to its
success probability $\widetilde m_\delta$ and rescale by $4F^2/\gamma$.
For an exact filter, $\widetilde m_\delta=\gamma E_\delta/(4F^2)$, where
$F=\norm{\bm{f}}$, so this directly recovers $E_\delta$ without preparing
the normalized solution state.
\Cref{app:numerical-resource-counts,app:numerical-energy-statistics}
specialize the query counts and error bound to this readout. As shown in
\cref{fig:numerical-query-comparison}(a) across sixteen complex source states,
coherent amplitude estimation outperforms independent Bernoulli sampling of
the same quantum success flag at equal controlled matrix-oracle budgets,
illustrating the $O(\epsilon_E^{-1})$ versus $O(\epsilon_E^{-2})$ precision
scaling~\cite{brassard2002quantum,montanaro2015quantum}. (On the $N=16$
circle benchmark in the numerical experiment, the large oracle
counts of $10^7$--$10^9$ reflect the high filter degree $d\le 255$ and tight
target precision $\epsilon_E\sim 10^{-4}$ rather than the system dimension $N$.)

To establish a rigorous quantum--classical comparison under identical access
to an unknown input, let $N$ be a power of two and suppose bits
$b_i\in\{0,1\}$ are queried via a reversible bit oracle. Defining
\begin{equation}
\label{eq:numerical-bit-family}
A=\operatorname{diag}(\sqrt{1+b_i}),\qquad
L=\operatorname{diag}(1+b_i),\qquad \bm{f}=N^{-1/2}\mathbf1,\qquad
E_\delta=1-\frac{1}{2N}\sum_{i=1}^N b_i,
\end{equation}
with all modes retained ($\gamma=1$, $\alpha_A=\sqrt2$), isolates scalar-energy
readout from manifold discretization. Preparing $\bm{f}$ via Hadamard gates and
implementing an exact cubic filter requires twelve bit-oracle calls per
energy-state preparation—all charged in \cref{fig:numerical-query-comparison}(b)—whereas
the randomized classical estimator uses one bit query per sample. Because
estimating $E_\delta$ on \eqref{eq:numerical-bit-family} reduces directly to
approximate counting, its worst-case fixed-confidence query complexity is
$\Theta(\epsilon_E^{-1})$ quantumly versus $\Theta(\epsilon_E^{-2})$ classically
whenever $N\gg\epsilon_E^{-2}$. The quantum bounds follow from
\cite{brassard2002quantum,nayak1999quantum}, and the matching randomized
classical lower bound is proved in \cref{app:numerical-bit-oracle}.

At the largest common budget of $44{,}040{,}108$ bit queries ($N=2^{36}$),
coherent estimation achieves a median 95th-percentile absolute energy error
of $6.99\times10^{-6}$, compared with $6.48\times10^{-5}$ for classical bit
sampling. Coherent readout outperforms classical sampling across all eleven
tested means, yielding a median paired error ratio of $8.90$ even after
charging the full twelve-query filter-preparation overhead.

Together, these benchmarks validate the implemented quadrature and
polynomial blocks and the prepared Galerkin state. They also demonstrate
substantial finite-budget precision gains under matched oracle access.
The conditional polylogarithmic dimension scaling is established
separately in \cref{thm:single-heat-estimation} and compared with
classical methods in \cref{tab:heat-dimension-comparison}.

\section{Summary}
\label{sec:summary}

We developed a quantum algorithmic framework connecting diffusion-map
manifold learning with discrete elliptic PDEs on point clouds through the
symmetric factorization $\mathcal G=A^TA$. Given coherent block-encoding
access to the factor $A$, Gaussian moment dilation, Laplace-weighted
unitary integrals, and QSVT evaluate heat-semigroup overlaps, regularized
inverse bilinear forms, and low-frequency Galerkin solution states. These
primitives support targeted learning and PDE queries without explicitly
assembling a dense kernel matrix or reconstructing a spectral basis.

Our complexity analysis quantifies the resulting reductions in both
operator and readout costs. At fixed normalized-overlap accuracy and
confidence, bounded $\alpha_A\sqrt t$, and polylogarithmic-cost coherent
access, heat estimation achieves polylogarithmic gate complexity in $N$,
compared with linear scaling for generic sparse explicit-vector methods.
For multiple queries, joint coherent estimation reduces shared heat-block
calls across $J$ overlaps from $\widetilde O(J/\epsilon)$ to
$\widetilde O(\sqrt J/\epsilon)$, with observable-specific operations
accounted for separately, while ideal circuit simulations validate our
heat and Galerkin constructions against their discrete targets.

More broadly, this work illustrates how machine learning and scientific
computing tasks can be formulated to align naturally with quantum
primitives. Selected predictions, regional averages, and low-frequency
spectral projections preserve the statistical and geometric quantities
needed for downstream inference while admitting direct evaluation via
matrix elements and operator functions. Incorporating input representations
and scalar readout requirements directly into the problem formulation
provides a principled route toward expanding the application layer of
fault-tolerant quantum computing.

Several established application domains highlight promising targets for
this spectral-filtering paradigm. In molecular simulation, diffusion maps
uncover collective reaction
coordinates~\cite{rohrdanz2011determination,ferguson2011nonlinear}, and
trajectory-based Galerkin methods approximate committors and mean
first-passage times~\cite{thiede2019galerkin}, motivating extensions to
elliptic boundary-value problems whose outputs are transition
probabilities or hitting-time averages. In climate science and nonlinear
dynamical systems, data-adapted Laplacian bases support the analysis of
intermittent dynamics and nonparametric
forecasting~\cite{giannakis2012nonlinear,berry2015nonparametric}, pointing
toward quantum representations of dynamical operators within learned
spectral spaces. Complementary targets arise in single-cell genomics,
where data diffusion denoises gene expression
measurements~\cite{vandijk2018recovering} and absorbing Markov chains on
diffusion graphs infer cell-fate
probabilities~\cite{setty2019characterization}, as well as in edge-aware
image editing~\cite{farbman2010diffusion} and graph signal
processing~\cite{shuman2013emerging}. Across these settings, smoothed
expression values, population averages, and diffusion similarities offer
well-defined scalar observables for application-level quantum benchmarks.

Realizing these applications end-to-end motivates several methodological
extensions. On the algorithmic and statistical side, key priorities
include constructing reusable factor and state-preparation oracles from
raw point clouds, jointly balancing kernel bandwidth, regularization,
spectral truncation, and quantum estimation accuracy against statistical
and discretization errors, and generalizing our routines to independently
learned Galerkin subspaces, nonhomogeneous boundary conditions, and
directed transition operators. In parallel, symmetric radial basis
function (RBF) Galerkin discretizations---widely used in geophysical
modeling and surface diffusion~\cite{fornberg2015primer,fuselier2013high,wendland1999meshless}---share
the same factored energy structure and offer a natural extension to
meshfree PDE solvers. Progress on these fronts will help bridge fault-tolerant quantum linear algebra with practical meshfree PDE and manifold-learning workflows. 

\bibliographystyle{plain}
\bibliography{ref}

\appendix
\section{Appendix}
\label{sec:appendix}

\subsection{Variational and Spectral Identities}
\label{app:variational-identities}
\subsubsection{Spectral Galerkin minimization and energy identities}
\label{app:galerkin-identities}
\begin{lemma}[Spectral Galerkin minimization]
\label{lem:appendix-galerkin-minimization}
Let $L=A^TA$, $\bm{f}\in\mathbb{R}^N\setminus\{0\}$, and $\delta>0$, with
$J$, $P_0$, $\Pi_\delta$, $P_\delta^+$, $\mathcal{E}_\delta$, and $B_\delta$
as in \cref{sec:setup-galerkin}. If $\bm{f}\perp\ker L$, then $\bm{u}=L^+\bm{f}$
is the unique solution of $L\bm{u}=\bm{f}$ and unique minimizer of $J$ in $(\ker L)^\perp$.
For every $\bm{f}$, the unique minimizer of $J$ on $\mathcal{E}_\delta$ is
\[
\bm{u}_\delta=B_\delta^+\bm{f}=L^+P_\delta^+\bm{f}
=\sum_{0<\lambda_j\leq\delta}\frac{\bm{q}_j^T\bm{f}}{\lambda_j}\bm{q}_j,
\qquad B_\delta^+=L^+\Pi_\delta,
\]
where $L\bm{q}_j=\lambda_j\bm{q}_j$ and the $\bm{q}_j$ form an orthonormal basis.
Equivalently, $\bm{u}_\delta\in\mathcal{E}_\delta$ satisfies
$L\bm{u}_\delta=P_\delta^+\bm{f}$, or the Galerkin condition
$\bm{v}^TL\bm{u}_\delta=\bm{v}^T\bm{f}$ for every $\bm{v}\in\mathcal{E}_\delta$. Moreover,
\[
E_\delta=\bm{f}^T\bm{u}_\delta=\bm{u}_\delta^TL\bm{u}_\delta=\norm{A\bm{u}_\delta}^2,
\qquad J(\bm{u}_\delta)=-\tfrac12E_\delta.
\]
If $P_\delta^+\bm{f}=0$, then $\bm{u}_\delta=0$ and $E_\delta=0$.
\end{lemma}
\begin{proof}
In the eigenbasis of $L$, its pseudoinverse inverts the positive eigenvalues,
and $B_\delta^+$ inverts only those in $(0,\delta]$. This gives the formulas,
$B_\delta^+=L^+\Pi_\delta=L^+P_\delta^+$, and both equivalent equations.
On either stated subspace, the corresponding solution $\bm{u}_*$ satisfies
\[
J(\bm{v})-J(\bm{u}_*)=\tfrac12(\bm{v}-\bm{u}_*)^TL(\bm{v}-\bm{u}_*),
\]
which is positive unless $\bm{v}=\bm{u}_*$ because $L$ is positive definite there.
The zero subspace has only the zero vector. Taking $\bm{v}=\bm{u}_\delta$ in the
Galerkin condition proves the energy identities. The spectral formula
gives the assertions when $P_\delta^+\bm{f}=0$.
\end{proof}

\begin{lemma}[Projected inverse, source overlap, and energy response]
\label{lem:appendix-galerkin-overlap-energy}
In the setting of \cref{lem:appendix-galerkin-minimization}, suppose
\cref{ass:galerkin-spectrum} holds, including its alternative when all
positive modes are retained. With $F=\norm{\bm{f}}$ and normalized $\ket{\hat f}$ from
\cref{ass:source-oracle}, $Q_\delta=(\gamma/4)L^+P_\delta^+$ satisfies
\[
0\preceq Q_\delta\preceq\tfrac14 I,\qquad
Q_\delta\ket{\hat f}=\frac{\gamma}{4F}\sum_{i=1}^N(\bm{u}_\delta)_i\ket{i},
\qquad s_\delta=\frac{\gamma\norm{\bm{u}_\delta}}{4F}.
\]
For $r_\delta=\norm{P_\delta^+\ket{\hat f}}$ as in
\cref{ass:galerkin-spectrum},
\[
\frac{\gamma r_\delta}{4\delta_-}\leq s_\delta\leq\frac{r_\delta}{4}.
\]
Thus a known $0<r_0\leq r_\delta$ gives the valid lower bound
$s_0=\gamma r_0/(4\delta_-)$ for solution-state preparation. When
$P_\delta^+\bm{f}\ne0$, $Q_\delta\ket{\hat f}/s_\delta=\ket{\hat u_\delta}$.
The energy response obeys
\[
m_\delta=\bra{\hat f}Q_\delta\ket{\hat f}
=\frac{\gamma E_\delta}{4F^2},\qquad
0\leq E_\delta\leq\frac{F^2}{\gamma}.
\]
These operator and energy identities also hold when $P_\delta^+\bm{f}=0$,
in which case $s_\delta=r_\delta=m_\delta=E_\delta=0$.
\end{lemma}
\begin{proof}
Every retained eigenvalue lies in $[\gamma,\delta_-]$, so $Q_\delta$
has eigenvalues $\gamma/(4\lambda_j)\in[\gamma/(4\delta_-),1/4]$
on $\mathcal{E}_\delta$ and zero elsewhere. Applying these bounds to
$P_\delta^+\ket{\hat f}$ proves the contraction and overlap bounds.
The remaining identities follow from $\bm{u}_\delta=L^+P_\delta^+\bm{f}$ and
$E_\delta=\bm{f}^T\bm{u}_\delta$. In particular, $0\leq m_\delta\leq1/4$
gives the energy bound, and the stated scalars vanish if $P_\delta^+\bm{f}=0$.
\end{proof}
\subsection{Supporting Operator Constructions}
\label{app:operator-constructions}

The dilation and compression identities follow from
\eqref{eq:dilation-square} by functional calculus, as in
\cite[Section~5.3]{wu2026quantum}.

\subsubsection{Gaussian heat-block implementation}
\label{app:gaussian-heat-cost-proof}

The Gaussian-LCHS implementation and its analysis are given in
\cite{kharazi2026sublinear}, with the factored-generator dilation also
developed in \cite[Section~5.3]{wu2026quantum}. We record only the
implementation scales needed for \cref{ass:gaussian-heat-cost}.
For $t>0$, set $H=B_A/\alpha_A$ and $\rho=\alpha_A\sqrt t$.
Truncate the Gaussian variable at $R=2\sqrt{\log(8/\eta)}$ and use
$2^n$ uniform radial cells of width $\Delta=R/2^n$, where
$n=\lceil\log_2\max\{2,16R(1+\rho)/\eta\}\rceil$.
The omitted mass and normalized midpoint error satisfy
\[
\epsilon_{\mathrm{tail+norm}}\leq\eta/4,
\qquad \epsilon_{\mathrm{quad}}\leq4(1+\rho)\Delta\leq\eta/4.
\]
At the midpoints $q_m=(m+\tfrac12)\Delta$, the Gaussian
amplitudes can be prepared by evaluating
$e^{-q_m^2/8}$ on a uniform register and coherently amplifying the
flag, whose success probability is $\Omega(1/R)$.
This uses $\widetilde O(\sqrt R)$ arithmetic and preparation steps
\cite{yoder2014fixed,an2026laplace}.
Binary time selection uses $O(n)$ controlled evolutions of $H$
with total duration below $\rho R$. Hamiltonian simulation therefore
costs $\widetilde O(n+\rho R)=\widetilde O(1+\alpha_A\sqrt t)$
matrix queries~\cite[full version, Theorem~58]{gilyen2019quantum}.
Preparation, arithmetic, and synthesis add only
polylogarithmic overhead. Assigning $\eta/8$ to preparation and
$\eta/4$ to selected evolution gives total error at most $\eta$
by \cref{lem:appendix-finite-construction-error}. Compression by $E_N$
and Hermitian symmetrization give the required Hermitian contraction
with unit block normalization. For $t=0$, use the identity.

\subsubsection{Normalization and Hermitian symmetrization of operator blocks}
\label{app:block-normalization}

The LCU and block-composition rules are proved in
\cite[full version, Lemmas~52--53]{gilyen2019quantum}.
Nonnegative weights summing to one give unit block normalization.
Ancilla-state and selected-evolution errors $\xi_{\mathrm P}$ and
$\xi_{\mathrm S}$ contribute at most $2\xi_{\mathrm P}+\xi_{\mathrm S}$,
including leakage from initialized simulation workspaces.
Compression by $E_N$ preserves this bound.

\begin{lemma}[Hermitian symmetrization]
\label{lem:appendix-hermitian-symmetrization}
If $U$ exactly block-encodes $X$ with normalization one, coherently
averaging $U$ and $U^\dagger$ with one selector ancilla block-encodes
$(X+X^\dagger)/2$ with the same normalization. For Hermitian $O$,
\[
\norm{\frac{X+X^\dagger}{2}}\leq1,\qquad
\norm{\frac{X+X^\dagger}{2}-O}\leq\norm{X-O}.
\]
\end{lemma}

\begin{lemma}[Normalization error]
\label{lem:appendix-normalization-error}
For a nonnegative combination or norm-convergent integral of unitaries
of total mass $s_\eta>0$, possibly followed by isometric compression,
\[
\norm{X/s_\eta-O}\leq\norm{X-O}+|s_\eta-1|.
\]
Indeed, $\norm{X}\leq s_\eta$ gives
$\norm{X/s_\eta-X}\leq|s_\eta-1|$.
\end{lemma}

\begin{lemma}[Error allocation for the finite constructions]
\label{lem:appendix-finite-construction-error}
For the Gaussian and Laplace constructions, omitted mass at most
$\eta/8$, normalized quadrature error $\eta/4$, ancilla-state error
$\eta/8$, and selected-evolution error $\eta/4$ give total error
\[
\eta/4+\eta/4+2(\eta/8)+\eta/4=\eta.
\]
Compression and Hermitian symmetrization yield a Hermitian contraction
with an exact unit-normalization block encoding. At
$\Lambda_{\beta,\eta}$ the Laplace tail is $\eta/8$.
The same symmetrization preserves the established Galerkin block error.
\end{lemma}

\subsection{Heat-semigroup Matrix-element Estimation}
\label{app:heat-estimation}

\subsubsection{Specialization for Theorem~\ref{thm:single-heat-estimation}}
\label{app:proof-single-heat-estimation}

\begin{proof}
Take $\eta=\epsilon/4$. The tagged state costs $O(T_u+T_v)$ to prepare,
and $\sigma_x\otimes\widetilde K(t)$ has a unit-normalization block
encoding of cost $O(T_K(t,\eta))$. By
\cref{lem:real-matrix-element-reduction}, its expectation differs from
$z(t)$ by at most $\eta$. Rall's estimator
\cite[Lemma~5]{rall2020quantum} contributes error $\epsilon/2$ with
failure probability $\delta_{\mathrm{fail}}$. Thus the total error is
$3\epsilon/4$, proving \eqref{eq:single-heat-guarantee}.
Substitution into \eqref{eq:rall-complexity} gives
\eqref{eq:single-heat-complexity}, including controls and inverses.
\end{proof}

\subsubsection{Specialization for Theorem~\ref{thm:joint-heat-estimation}}
\label{app:proof-joint-heat-estimation}

\begin{proof}
Set $\eta=\epsilon/4$ and $\nu=\epsilon/2$. Each preparation or inverse
of \eqref{eq:heat-common-state} costs $O(T_v+T_K(t,\eta))$ and uses
one controlled heat-block call. The construction in
\cref{eq:heat-query-observables,eq:heat-common-expectation} gives norm-one
Hermitian observables with uniform bias at most $\eta$ from the real
targets $m_i(t)$. Their controlled evolutions in
\eqref{eq:heat-observable-evolution} cost
$O(T_{u_i}+T_{\mathrm{rot}})$ and use no $U_v$ or $U_K$.

Apply \cite[Theorem~1]{huggins2022multiple} with error $\nu$.
Its $\widetilde O(\sqrt J/\nu)$ common-state calls, repeated
$O(\log(1/\delta_{\mathrm{fail}}))$ times with coordinatewise medians,
give simultaneous error at most $\nu$ with the requested confidence.
The total error is $\eta+\nu=3\epsilon/4$, proving
\eqref{eq:joint-heat-guarantee}. Multiplication by the common-state
preparation cost gives \eqref{eq:joint-heat-common-cost}.
Observable-processing and classical-output costs are additional.
The complete common state is retained, so no postselection factor enters.
\end{proof}

For total gate costs, let $C_t=T_v+T_K(t,\epsilon/4)$ and
$S_u=\sum_i(T_{u_i}+T_{\mathrm{rot}})$, including controlled rotations
and synthesis. The same estimator gives
\begin{equation}
\label{eq:joint-heat-total-cost}
T_{\mathrm{joint}}^{(J)}
=\widetilde O\!\left(\frac{\sqrt J}{\epsilon}(C_t+S_u)
\log\frac1{\delta_{\mathrm{fail}}}\right).
\end{equation}
Independent sampling and coherent estimation give, respectively,
\begin{equation}
\label{eq:heat-independent-readout-costs}
\begin{aligned}
T_{\mathrm{samp}}^{(J)}
&=\widetilde O\!\left(\frac{JC_t+S_u}{\epsilon^2}
\log\frac{J}{\delta_{\mathrm{fail}}}\right),\\
T_{\mathrm{ind}}^{(J)}
&=\widetilde O\!\left(\frac{JC_t+S_u}{\epsilon}
\log\frac{J}{\delta_{\mathrm{fail}}}\right).
\end{aligned}
\end{equation}
Classical processing and output costs are additional.

\subsection{Regularized Inverse Scalar Estimation}
\label{app:inverse-estimation}

\subsubsection{Resolvent representation and truncation}
\label{app:resolvent-representation}

\begin{lemma}[Resolvent representation and truncation]
\label{lem:appendix-resolvent-representation}
For every $\beta>0$, the normalized resolvent in
\eqref{eq:normalized-resolvent} satisfies $0\preceq R_\beta\preceq I$ and
\begin{equation*}
R_\beta=\int_0^\infty\beta e^{-\beta t}K(t)\,dt
=E_N^T\!\left(\int_{\mathbb{R}}p_\beta(k)e^{-ikB_A}\,dk\right)E_N.
\end{equation*}
Both integrals converge absolutely in operator norm, and $p_\beta$ is a
probability density. For every $\Lambda>0$, the definitions in
\cref{eq:resolvent-tail,eq:resolvent-cutoff} obey
\begin{equation*}
\begin{aligned}
\norm{R_\beta-R_{\beta,\Lambda}}&\leq e^{-\sqrt{\beta}\Lambda},\\
\norm{E_N^T\!\left(\int_{-\Lambda}^{\Lambda}
\bar p_{\beta,\Lambda}(k)e^{-ikB_A}\,dk\right)E_N-R_\beta}
&\leq2e^{-\sqrt{\beta}\Lambda}.
\end{aligned}
\end{equation*}
For $0<\eta<1/4$, $\Lambda=\Lambda_{\beta,\eta}$ makes the combined
truncation and normalization error at most $\eta/4$.
\end{lemma}

\begin{proof}
The scalar identities
\[
\int_0^\infty\beta e^{-(\beta+\lambda)t}\,dt
=\frac{\beta}{\beta+\lambda},\qquad
\int_{\mathbb R}p_\beta(k)e^{-ikx}\,dk
=\frac{\beta}{\beta+x^2}
\]
give the representations by spectral calculus and
\eqref{eq:dilation-square}, as in the Laplace-transform
framework~\cite{an2026laplace}. The omitted probability is
$e^{-\sqrt\beta\Lambda}$. Unitarity bounds truncation by this mass,
and \cref{lem:appendix-normalization-error} gives the same bound for
renormalization. At $\Lambda_{\beta,\eta}$ their sum is $\eta/4$.
\end{proof}

\subsubsection{Coherent Laplace quadrature}
\label{app:explicit-laplace-quadrature}

The quadrature and coherent LCU analysis follows
\cite[Sections~3.2--3.5 and Appendix~B]{an2026laplace}.
We record the geometric cell weights for the density in
\eqref{eq:square-root-resolvent}, which give its preparation and cost
under the factor-access model.

\begin{lemma}[Geometric-weight Laplace quadrature]
\label{lem:appendix-explicit-laplace-quadrature}
\label{ass:inverse-quadrature}
Let $\beta,\Lambda,\alpha_A>0$, $\norm{B_A}\leq\alpha_A$, and $0<\zeta<1/4$.
Set
\begin{equation}
\label{eq:appendix-laplace-grid}
\begin{gathered}
n=\left\lceil\log_2\max\left\{2,
\frac{\alpha_A\Lambda}{2\zeta}\right\}\right\rceil,
\quad M_{\mathrm q}=2^n,\quad \Delta=\Lambda/M_{\mathrm q},
\quad r=e^{-\sqrt\beta\Delta},\\
k_{s,m}=(-1)^s(m+\tfrac12)\Delta,\qquad
w_m=\frac{(1-r)r^m}{1-r^{M_{\mathrm q}}},
\quad s\in\{0,1\},\quad 0\leq m<M_{\mathrm q}.
\end{gathered}
\end{equation}
The normalized weights $w_m/2$ satisfy
\begin{equation}
\label{eq:appendix-laplace-midpoint-error}
\left\|\frac12\sum_{s,m}w_me^{-ik_{s,m}B_A}
-\int_{-\Lambda}^{\Lambda}
\bar p_{\beta,\Lambda}(k)e^{-ikB_A}\,dk\right\|
\leq\alpha_A\Delta/2\leq\zeta.
\end{equation}
Their preparation state factors as
\begin{equation}
\label{eq:appendix-laplace-product-state}
\ket+\otimes\sum_m\sqrt{w_m}\ket m
=\ket+\otimes\bigotimes_{j=0}^{n-1}
\frac{\ket0+r^{2^j/2}\ket1}{\sqrt{1+r^{2^j}}}.
\end{equation}
Binary time selection uses $O(n)$ controlled evolutions with total
duration below $\Lambda$. For preparation error $\xi_{\mathrm P}$
and selected-evolution error $\xi_{\mathrm S}$, the matrix-query
and gate costs, including synthesis, are
\begin{equation}
\label{eq:appendix-laplace-select-cost}
\widetilde O(n+\alpha_A\Lambda),\qquad
\widetilde O\bigl((n+\alpha_A\Lambda)T_A+n\bigr).
\end{equation}
Logarithmic factors in $1/\xi_{\mathrm P}$ and $1/\xi_{\mathrm S}$
are suppressed.
\end{lemma}
The error bound follows from
$\|e^{-ikB_A}-e^{-ik'B_A}\|\leq\alpha_A|k-k'|$ within each cell.
The product identity follows from
$\prod_{j=0}^{n-1}(1+r^{2^j})=(1-r^{2^n})/(1-r)$.
Thus only $n$ one-qubit rotations and one sign Hadamard are needed
before synthesis. The remaining implementation follows the cited
LCU framework, with preparation and selection errors contributing
$2\xi_{\mathrm P}+\xi_{\mathrm S}$.

\subsubsection{Specialization for Theorem~\ref{thm:inverse-block-encoding}}
\label{app:proof-inverse-block-encoding}

\begin{proof}
At $\Lambda=\Lambda_{\beta,\eta}$, apply
\cref{lem:appendix-explicit-laplace-quadrature} with
$\zeta=\eta/4$, $\xi_{\mathrm P}=\eta/8$, and $\xi_{\mathrm S}=\eta/4$.
The total error, including truncation and normalization, is
$\eta/4+\zeta+2\xi_{\mathrm P}+\xi_{\mathrm S}=\eta$.
Compression and Hermitian symmetrization give the contraction in
\eqref{eq:inverse-block-accuracy}. The quadrature has
$n=O(\log(2+\alpha_A\Lambda_{\beta,\eta}/\eta))$ and total evolution
time $O(\Lambda_{\beta,\eta})$, hence
\[
n_{B,R}(\beta,\eta)=\widetilde O(1+\alpha_A/\sqrt\beta),\qquad
T_R(\beta,\eta)=\widetilde O\!\left((1+\alpha_A/\sqrt\beta)T_A\right).
\]
This proves \eqref{eq:inverse-block-cost}, including preparation,
controls, inverses, compression, symmetrization, and synthesis.
\end{proof}

\subsubsection{Specialization for Theorem~\ref{thm:inverse-scalar-estimation}}
\label{app:proof-inverse-scalar-estimation}

\begin{proof}
The bound $|r_\beta(\bm{y},\bm{b})|\leq1/\beta$ proves the zero-output branch
when $\beta\epsilon\geq1$. Otherwise apply
\cref{app:proof-single-heat-estimation} with $U_K$ replaced by $U_R$,
target $\beta r_\beta(\bm{y},\bm{b})$, $\eta=\beta\epsilon/4$, and
$\nu=\beta\epsilon/2$. The tagged identity
\eqref{eq:inverse-tagged-expectation} has bias at most $\eta$.
Rall's estimator~\cite[Lemma~5]{rall2020quantum} and division by $\beta$
give, with probability at least $1-\delta_{\mathrm{fail}}$,
\[
|\widehat r_\beta(\bm{y},\bm{b})-r_\beta(\bm{y},\bm{b})|
\leq\frac{\nu+\eta}{\beta}=\frac{3\epsilon}{4}<\epsilon,
\]
proving \eqref{eq:inverse-scalar-guarantee}. Substituting
\eqref{eq:inverse-block-cost} into the same cost calculation gives
\eqref{eq:inverse-scalar-cost}. As in the heat proof, no postselection
or factor $\norm{R_\beta\ket{b}}^{-1}$ is required.
\end{proof}

\subsubsection{Specialization for Theorem~\ref{thm:joint-inverse-estimation}}
\label{app:proof-joint-inverse-estimation}

\begin{proof}
When $\beta\epsilon\geq1$, return zero. Otherwise repeat
\cref{app:proof-joint-heat-estimation} with
$(U_K,\ket v,\ket{u_i})$ replaced by $(U_R,\ket b,\ket{y_i})$,
$\eta=\beta\epsilon/4$, and $\nu=\beta\epsilon/2$.
The uniform bias and simultaneous estimation error sum to
$3\beta\epsilon/4$. Dividing by $\beta$ proves
\eqref{eq:joint-inverse-guarantee}.
Each common-state call costs $O(T_b+T_R(\beta,\eta))$, and each
observable evolution costs $O(T_{y_i}+T_{\mathrm{rot}})$ with no
$U_b$ or $U_R$ calls. Substitution of \eqref{eq:inverse-block-cost}
gives \cref{eq:joint-inverse-common-cost,eq:joint-inverse-total-cost}.
\end{proof}

\subsubsection{Operator-scale comparison and complexity specializations}
\label{app:inverse-complexity-specializations}

Multiplying two matrix blocks with independent signal ancillas and
compressing gives a block encoding of $\mathcal G$ with normalization
$\alpha_A^2$, since
$E_N^T(B_A/\alpha_A)^2E_N=\mathcal G/\alpha_A^2$.
It uses $O(1)$ calls to $U_B$ and gate cost $\widetilde O(T_A)$.
The Laplace tail beyond $t=\beta^{-1}\ln(1/\eta)$ has norm at most
$\eta$. Under the coherent quadrature assumptions of direct
Lap-LCHS~\cite{an2026laplace}, its Fourier cutoff contributes only
polylogarithmic factors, giving matrix-oracle cost
$\widetilde O(1+\alpha_A^2/\beta)$ for a constant-normalized resolvent
block. In contrast, \cref{thm:inverse-block-encoding} has scale
$\widetilde O(1+\alpha_A/\sqrt\beta)$.
For $0<\beta\epsilon<1$, applying the same observable estimator to the
direct construction gives
\begin{equation*}
\widetilde O\!\left(
\frac{1+\alpha_A^2/\beta}{\beta\epsilon}
\log\frac1{\delta_{\mathrm{fail}}}\right)
\end{equation*}
matrix-oracle calls, to compare with \eqref{eq:inverse-scalar-cost}.
That full bound also immediately yields
\begin{equation*}
n_{B,\mathrm{scalar},R}=\begin{cases}
\widetilde O\!\left(\dfrac{\alpha_A}{\beta^{3/2}\epsilon}
\log\dfrac1{\delta_{\mathrm{fail}}}\right),
&\alpha_A/\sqrt\beta\geq1,\\[2mm]
\widetilde O\!\left(\epsilon^{-1}
\log\dfrac1{\delta_{\mathrm{fail}}}\right),
&\alpha_A=O(1),\quad\beta=\Theta(1).
\end{cases}
\end{equation*}
The second line is the normalized graph Laplacian specialization with
its assumed block access. State-preparation costs remain those in
\eqref{eq:inverse-scalar-cost}, and the zero-output branch costs no
oracle calls when $\beta\epsilon\geq1$.
\subsection{Spectral Galerkin State Preparation}
\label{app:galerkin-estimation}

\subsubsection{Projected-inverse construction}
\label{app:proof-galerkin-projected-inverse}

For $0<\eta_Q<1$, set $\eta_p=\eta_Q/4$,
$a_\gamma=\sqrt\gamma/\alpha_A$, and $b_\pm=\sqrt{\delta_\pm}/\alpha_A$.
The reciprocal and cutoff approximations
\cite[full version, Corollary~67 and Lemma~29]{gilyen2019quantum}
give bounded polynomials $r,h$, with $r$ odd and $h$ even.
Choose $r$ to approximate $a_\gamma/(2x)$ on $a_\gamma\leq|x|\leq1$,
and $h$ to approximate one on $|x|\leq b_-$ and zero on $|x|\geq b_+$,
each to error $\eta_p/2$. Then $p=rh$ satisfies
\eqref{eq:galerkin-reciprocal-polynomial}, with degree
$\widetilde O(a_\gamma^{-1}+(b_+-b_-)^{-1})$.
Take $h=1$ and omit $b_+$ when all positive modes are retained.
The squared polynomial
$q=p^2$ is even, bounded by one on $[-1,1]$, and vanishes at zero.
Its degree is $2\deg p=d_Q(\eta_Q)$, as in
\eqref{eq:galerkin-qsvt-degree}.

For a retained eigenvalue, write
$a_j=\sqrt\gamma/(2\sqrt{\lambda_j})\leq1/2$. Then
\begin{equation*}
\left|p(\sqrt{\lambda_j}/\alpha_A)^2-a_j^2\right|
\leq\eta_p(2a_j+\eta_p)\leq\eta_p+\eta_p^2.
\end{equation*}
On discarded modes the squared polynomial is at most $\eta_p^2$, and
both operators vanish on the nullspace. Spectral compression therefore gives
\begin{equation*}
\norm{Q_{\delta,p}-Q_\delta}\leq\eta_p+\eta_p^2.
\end{equation*}

Real-polynomial QSVT~\cite[full version, Corollary~18]{gilyen2019quantum}
implements this transform with unit block normalization. Allocate
$\eta_Q/2$ to implementation error and, when needed, apply
\cref{lem:appendix-hermitian-symmetrization}. Compression and symmetrization
preserve the normalization and do not increase the error, so the resulting
exact Hermitian contraction satisfies
\begin{equation*}
\norm{\widetilde Q_\delta-Q_\delta}
\leq\eta_p+\eta_p^2+\frac{\eta_Q}{2}
=\frac{3\eta_Q}{4}+\frac{\eta_Q^2}{16}<\eta_Q.
\end{equation*}
This proves \eqref{eq:galerkin-implemented-block}.

The implementation uses $O(d_Q(\eta_Q))$ matrix-oracle calls and signal
operations, giving \eqref{eq:galerkin-block-cost}. Embedding, synthesis,
and controls are charged as specified in \cref{ass:galerkin-spectrum}.
When all positive modes are retained, take $h=1$ and omit the
$\Delta_\sigma^{-1}$ contribution.

\subsubsection{Specialization for Theorem~\ref{thm:galerkin-state}}
\label{app:proof-galerkin-state}

\begin{proof}
Set $\eta_Q=\epsilon_{\mathrm{st}}s_0/8$ and
$\bm{w}=\widetilde Q_\delta\ket{\hat f}$. By
\eqref{eq:galerkin-implemented-block} and
\cref{lem:appendix-galerkin-overlap-energy},
\[
\norm{\bm{w}}\geq s_\delta-\eta_Q\geq\frac78s_0>\frac12s_0,
\qquad
\norm{\frac{\bm{w}}{\norm{\bm{w}}}-\ket{\hat u_\delta}}
\leq\frac{2\eta_Q}{s_\delta}\leq\frac{\epsilon_{\mathrm{st}}}{4}.
\]
Fixed-point amplification~\cite{yoder2014fixed}, with amplitude bound $s_0/2$,
preserves the successful direction up to a global phase and achieves ideal
failure probability $\delta_{\mathrm{fail}}/2$ using
$O(s_0^{-1}\log(2/\delta_{\mathrm{fail}}))$ calls to $W_Q,W_Q^\dagger$.

The synthesis error is
$\xi\leq\min\{\epsilon_{\mathrm{st}}/16,\delta_{\mathrm{fail}}/8\}$.
It changes success probability by at most $2\xi$. Since the ideal successful
component has norm at least $1/\sqrt2$, normalization changes its direction
by at most $2\sqrt2\,\xi$. Thus
\[
p_{\mathrm{succ}}\geq1-\frac{\delta_{\mathrm{fail}}}{2}
-2\xi\geq1-\delta_{\mathrm{fail}},\qquad
d_{\mathrm{ph}}(\ket{\psi_\delta},\ket{\hat u_\delta})
\leq\frac{\epsilon_{\mathrm{st}}}{4}
+\frac{\sqrt2\,\epsilon_{\mathrm{st}}}{8}<\epsilon_{\mathrm{st}}.
\]
This proves \eqref{eq:galerkin-state-guarantee}. Each amplification step
uses $O(d_Q(\eta_Q))$ matrix-oracle calls, $O(1)$ source-oracle calls,
and $O(d_Q(\eta_Q)(T_A+T_{\mathrm{sig}})+T_f+T_{\mathrm{ref}})$ gates.
Multiplying by the amplification count gives
\cref{eq:galerkin-state-queries,eq:galerkin-state-cost}.
The alternative choice $s_0=\gamma r_0/(4\delta_-)$ follows from
\cref{lem:appendix-galerkin-overlap-energy}.
\end{proof}

\subsection{Analysis Supporting the Numerical Comparisons}
\label{app:numerical-circuits}

We retain the cost and error arguments used in the main text.

\subsubsection{Classical heat-action comparison}
\label{app:heat-classical-comparison}

For normalized inputs, vector-action error $\epsilon$ bounds the error
in $\bra{u}K(t)\ket{v}$ by $\epsilon$.
A Chebyshev approximation on $[0,\norm{\mathcal G}]$ has sufficient degree
\[
k=O\!\left(\sqrt{t\norm{\mathcal G}\log(2/\epsilon)}
          +\log(2/\epsilon)\right)
\]
\cite[Theorem~4.1]{sachdeva2014faster}.
Its recurrence and final inner product cost
$O(k(C_{\mathcal G}+N))$, where
$C_{\mathcal G}=O(\operatorname{nnz}(\mathcal G)+N)$ for a sparse matrix
or $O(\operatorname{nnz}(A)+M+N)$ through $A^T(A\bm{x})$.
The same form applies to scaled Taylor action, with its own step count
\cite{almohy2011action}. Lanczos adds an $O(k^2)$ tridiagonal eigensolve,
while full reorthogonalization adds $O(Nk^2)$ work
\cite{hochbruck1997krylov,guettel2023sketching}.
Rational Krylov costs also include shifted solves
\cite{guettel2013rational}. Circulant Fourier evaluation costs
$O(N\log N)$ for generic explicit inputs.

These are costs of explicit-vector algorithms. For
$\mathcal G=I-P$ with symmetric stochastic $P$,
$K(t)=e^{-t}\sum_{j\geq0}t^jP^j/j!$.
Neighbor sampling therefore estimates one entry using
$O(\epsilon^{-2})$ Poisson-length walks of mean $t$, with expected
$O((1+t)\epsilon^{-2})$ local work at fixed confidence.
Local diffusion methods also have graph-size-independent guarantees
for suitable norms and seeds~\cite{kloster2014heat}.

\subsubsection{Heat-circuit validation}
\label{app:numerical-heat-validation}

The Gaussian-LCU tests compare the implemented block with its quadrature
target and compare that target with the matrix exponential. The
four-node test separates the $6.12\times10^{-16}$ circuit discrepancy
from the $2.83154\times10^{-3}$ quadrature bias. The accompanying
numerical experiment gives the quadrature sweep and compiled circuits.
These Gauss--Hermite tests validate the representation used in
\cref{sec:numerical-circle}. The uniform-grid implementation summarized
in \cref{app:gaussian-heat-cost-proof} supplies the asymptotic cost in
\cref{ass:gaussian-heat-cost}.

\subsubsection{Circle and exact-fixture checks}
\label{app:numerical-example-identities}

For the circle discretization in \eqref{eq:numerical-circle-coordinates},
$\lambda_k=4h^{-2}\sin^2(\pi k/N)$ and the retained forcing gives
\[
z_{\delta,i}=\frac{\sin\theta_i}{\lambda_1}
+\frac{\cos(2\theta_i)}{\lambda_2},\qquad
E_\delta=\pi\left(\frac1{\lambda_1}+\frac1{\lambda_2}\right).
\]
The expansion $\lambda_k=k^2-k^4h^2/12+O(h^4)$ gives the reported
$O(h^2)$ nodal error. The circuit target is the discrete
solution, so circuit error and continuum discretization error are
assessed separately.

\subsubsection{Supplementary energy readout}

For the benchmarks in \cref{sec:numerical-query-advantage}, the projected inverse also gives
\begin{equation}
\label{eq:galerkin-energy}
E_\delta=\bm{f}^T\bm{u}_\delta=\bm{u}_\delta^TL\bm{u}_\delta=\norm{A\bm{u}_\delta}^2,
\qquad J(\bm{u}_\delta)=-\tfrac12E_\delta,
\end{equation}
and, for known $F=\norm{\bm{f}}$,
\begin{equation}
\label{eq:galerkin-energy-expectation}
m_\delta:=\bra{\hat{f}}Q_\delta\ket{\hat{f}}
=\frac{\gamma E_\delta}{4F^2},\qquad
0\leq E_\delta\leq\frac{F^2}{\gamma}.
\end{equation}
These identities permit scalar readout without preparing the normalized
solution state. \Cref{alg:galerkin-energy} follows Rall's observable
estimator~\cite[Lemma~5]{rall2020quantum}. The numerical benchmarks in
\cref{sec:numerical-query-advantage} use canonical amplitude estimation
of the one-filter success probability, with the same output rescaling.

\begin{algorithm}[H]
\caption{Supplementary spectral Galerkin energy readout}
\label{alg:galerkin-energy}
\begin{algorithmic}[1]
\Require Oracles $U_B,U_f$, the spectral parameters in
\cref{ass:galerkin-spectrum}, known $F$, $\epsilon_E>0$, and
$0<\delta_{\mathrm{fail}}<1/2$.
\If{$\epsilon_E\geq F^2/\gamma$}
\State \Return $0$.
\EndIf
\State Construct $U_Q$ with $\eta_Q=\gamma\epsilon_E/(16F^2)$.
\State Estimate $m_\delta$ using $\widetilde Q_\delta$ in
$\ket{\hat f}$ to accuracy $\nu_E=\gamma\epsilon_E/(8F^2)$
and failure probability $\delta_{\mathrm{fail}}$.
\State \Return
$\operatorname{clip}_{[0,F^2/\gamma]}(4F^2\widehat m_\delta/\gamma)$.
\end{algorithmic}
\end{algorithm}
The rescaled block and estimation errors sum to $3\epsilon_E/4$.
The one-filter implementation and its charged query counts are given
in \cref{app:numerical-resource-counts}.

\subsubsection{Energy readout and charged queries}
\label{app:numerical-resource-counts}

The benchmarks use a single odd filter rather than the squared filter
used for solution-state preparation. Since odd QSVT maps the input
sector to the output sector, its success probability is
\begin{equation}
\label{eq:numerical-one-filter-energy}
\widetilde m_\delta
=\|p^{(\mathrm{SV})}(A/\alpha_A)\ket{\hat f}\|^2
=\bra{\hat f}Q_{\delta,p}\ket{\hat f},\qquad
E_{\delta,p}=\frac{4F^2}{\gamma}\widetilde m_\delta.
\end{equation}
The implemented real-polynomial selector uses $2d$ controlled
matrix-oracle calls for $d=\deg p$, including its inverse sequence.
With $\ell$ phase qubits and $R$ repetitions, canonical amplitude
estimation~\cite{brassard2002quantum} therefore uses
\begin{equation}
\label{eq:numerical-qae-counts}
\begin{aligned}
\text{matrix-oracle calls}&=2dR(2^{\ell+1}-1),\\
\text{source-oracle calls}&=R(2^{\ell+1}-1).
\end{aligned}
\end{equation}
For the bit family in \eqref{eq:numerical-bit-family}, use the exact
cubic filter $p(x)=(3x-2x^3)/(2\sqrt2)$. Each matrix query costs two
bit queries, giving $Q_b=12R(2^{\ell+1}-1)$.
These counts are for the benchmark circuits.

\subsubsection{Energy accuracy and the classical lower bound}
\label{app:numerical-energy-statistics}
\label{app:numerical-bit-oracle}

We use the success-probability error bound proved for canonical
amplitude estimation in~\cite{brassard2002quantum}:
\[
|\widehat m_\delta-\widetilde m_\delta|
\leq\frac{2\pi\sqrt{\widetilde m_\delta(1-\widetilde m_\delta)}}{2^\ell}
+\frac{\pi^2}{2^{2\ell}}
\]
with probability at least $8/\pi^2$. Taking the median of seven
independent runs gives failure probability at most
$\sum_{k=4}^7\binom7k(1-8/\pi^2)^k(8/\pi^2)^{7-k}<0.02764$.
At fixed confidence and operator scales, this gives the
$O(\epsilon_E^{-1})$ coherent-estimation cost, compared with
$O(\epsilon_E^{-2})$ independent sampling of the same success flag.
The error relative to the intended energy also includes filter bias:
\begin{equation}
\label{eq:numerical-energy-error-budget}
|\widehat E_\delta-E_\delta|
\leq|\widehat E_\delta-E_{\delta,p}|
+\frac{4F^2}{\gamma}\|Q_{\delta,p}-Q_\delta\|
\leq|\widehat E_\delta-E_{\delta,p}|
+\frac{4F^2}{\gamma}(\eta_p+\eta_p^2).
\end{equation}
Circuit implementation error is added to the block discrepancy.

For the matched bit-oracle family in \eqref{eq:numerical-bit-family},
the energy is one minus half the bit mean.
For the classical lower bound against adaptive randomized queries, take
$0<\epsilon_E<1/16$ and independent Bernoulli bits with biases
\begin{equation*}
q_-=\frac12-4\epsilon_E,\qquad q_+=\frac12+4\epsilon_E.
\end{equation*}
For sufficiently large $N\epsilon_E^2$, Hoeffding's inequality gives
$|N^{-1}\sum_i b_i-q_\pm|\leq\epsilon_E$ with probability at least
$0.99$. Energy error $\epsilon_E$ gives bit-mean error $2\epsilon_E$
through $2(1-\widehat E_\delta)$. An estimator succeeding with probability
at least $2/3$ on every input therefore distinguishes the two hypotheses
with constant advantage by comparing its bit-mean estimate with $1/2$.

Conditional on the transcript, each unqueried bit remains
$\operatorname{Bernoulli}(q_\pm)$ and contributes $O(\epsilon_E^2)$
relative entropy. Repeated queries add no information. Including the
independent random seed, the chain rule bounds transcript relative
entropy by $O(Q\epsilon_E^2)$ after $Q$ adaptive queries. Pinsker's
inequality requires constant relative entropy for constant distinguishing
advantage, giving $Q=\Omega(\epsilon_E^{-2})$. Uniform sampling matches
this bound. Coherent estimation has optimal fixed-confidence complexity
$\Theta(\epsilon_E^{-1})$ in this regime
\cite{brassard2002quantum,nayak1999quantum,montanaro2015quantum}.
Exact classical enumeration costs $N$ queries and supersedes sampling
when cheaper. Direct quantum bit-mean estimation
has smaller constants than this benchmark, which charges the full
twelve-query QSP preparation.

\end{document}